\documentclass[twocolumn]{autart}
\usepackage[utf8]{inputenc}
\usepackage{amsmath}
\usepackage{amsfonts}
\usepackage{amssymb}
\usepackage{graphicx}
\usepackage[dvipsnames]{xcolor}
\usepackage{booktabs}
\usepackage{epstopdf}
\usepackage{mathrsfs}
\usepackage{subfigure}
\usepackage{tabulary}

\usepackage{savesym}
\savesymbol{AND}
\usepackage{algorithm}
\usepackage{algcompatible}
\usepackage{calc}

\newcommand*{\LongState}[1]{\State
\parbox[t]{\linewidth-\algorithmicindent-\algorithmicindent}{#1\strut}}

\newcommand{\E}{{\mathbb E}}
\newcommand{\Z}{{\mathbb Z}}
\newcommand{\R}{{\mathbb R}}

\newcommand{\N}{{\mathbb N}}

\newtheorem{lemmer}{Lemma}[section]
\newtheorem{remark}[lemmer]{Remark}

\begin{document}

\begin{frontmatter}

%

\thanks{ This work has been partially supported by the FIRB project ``Learning meets time'' (RBFR12M3AC) funded by MIUR and by Progetto di Ateneo CPDA147754/14 ``New statistical learning approach for multi-agents adaptive estimation and coverage control''.}
\title{Maximum Entropy Vector Kernels for MIMO system identification}

\author[]{Giulia Prando}\ead{prandogi@dei.unipd.it},
\author[]{Gianluigi Pillonetto}\ead{giapi@dei.unipd.it},
\author[]{Alessandro Chiuso}\ead{chiuso@dei.unipd.it}

\address{
    Department of Information Engineering \\
    University of Padova, Padova (Italy)
}

\begin{abstract}
Recent contributions have framed   linear system identification  as a nonparametric regularized inverse problem. {\color{black} Relying on  $\ell_2$-type regularization which accounts for the stability and smoothness of the impulse response to be estimated, these approaches  have been shown to be competitive   w.r.t classical parametric methods. In this paper,  adopting  Maximum Entropy arguments, we  derive a new $\ell_2$ penalty deriving from  a vector-valued kernel; to do so we exploit the structure of the Hankel matrix, thus controlling at the same time  complexity, measured by  the McMillan degree,  stability and  smoothness of the identified models.}  As a special case we recover the nuclear norm penalty on the squared block Hankel matrix. 
In contrast with previous literature on reweighted nuclear norm penalties, our  kernel is described by a small number of hyper-parameters, which are iteratively updated through marginal likelihood maximization; constraining the structure of the kernel  acts as a (hyper)regularizer which helps controlling the effective degrees of freedom of our estimator. To optimize the marginal likelihood we adapt a Scaled Gradient Projection (SGP) algorithm  which is proved to be significantly computationally cheaper than other first and second order off-the-shelf optimization methods. The paper also contains an extensive comparison with many state-of-the-art methods on several Monte-Carlo studies, which confirms the effectiveness of our procedure.  
\end{abstract}



\end{frontmatter}

\section{Introduction} \label{sec:intro}

\noindent Although linear system identification is  sometimes considered a mature field, with a wide and solid literature summarized in the well known textbooks \cite{Ljung:99,Soderstrom}, the recent developments on regularization based methods have brought new insights and opened new avenues. The most common ``classical'' approaches are parametric Prediction Error Methods (PEM)  \cite{Ljung:99,Soderstrom}, where model classes  (OE, ARMAX, Box-Jenkins, state-space, etc.) are described by a finite dimensional parameter vector which is estimated minimizing the squared prediction errors, and subspace methods, which translate ideas from stochastic  realization theory \cite{Anders_birthday,LPbook} into algorithms which work on measured data \cite{Vanov-book}. 

{\color{black}These techniques require that a model complexity (the \emph{order} hereon) is fixed, and thus estimated, first. 
As an alternative} to the standard parametric approach, recent literature has proposed  a Bayesian perspective, leading to a class of regularized methods \cite{SS2010,SS2011,ChenOL12,SurveyKBsysid,ChenetalTAC2014,BSLCDC,BSL_JOURNAL}. The use of Bayesian inference is not new in the field of identification and time-series estimation: early works on this topic appeared in the late'70, early '80 \cite{Akaike1979,DoanLSER1984,KitagawaTAC1985,Goodwin1992}; see \cite{ChiusoARC2016} for an  overview.\\
The Bayesian paradigm considers the impulse response as a stochastic process whose prior distribution penalizes undesired systems (e.g. unstable ones). 
This allows to face the so-called bias/variance trade-off by jointly performing estimation and model  selection.


\vspace{3mm}

\noindent In \cite{SS2010,SS2011,ChenOL12}  prior distributions  are designed to encode  smoothness and stability of the impulse response to be estimated, leading to  $\ell_2$-type penalties so that closed-form solution are available. These priors can also be shown to be solutions of  Maximum Entropy problems, see \cite{PillonettoN11,NicFer:1998:IFA_1779,CarliCL2015,ChenACCLP15}. 
\\In this paper, we focus on the identification of multi input-multi output (MIMO) systems, where matrix impulse responses  have to be  identified. Similar problems are encountered in multi-task learning where one would like to simultaneously estimate multiple functions while also exploiting their mutual information. To this aim \cite{Caruana:1997,Bakker:2003,Micchelli:2005,Evgeniou:2005,PillPAMI} have considered vector-valued kernels which account for the smoothness of the functions to be estimated. In the identification of finite dimensional linear MIMO systems, the coupling between different input-output channels is captured  by Hankel matrix, which has finite rank equal to the McMillan degree of the system.
\\The Hankel matrix and its properties have  already been thoroughly exploited in subspace methods, where also Vector AutoRegressive Models (VARX) estimated under the PEM framework play a fundamental role; in fact it has been shown in \cite{ChiusoAUTO06} (see also \cite{ChiusoTAC2007,ChiusoTAC2010}) that certain subspace methods can be seen as estimation of a long (i.e. ``nonparametric'' in the context of this paper) VARX model followed by a suitable (data based) model reduction step. This paper goes one step further, by  merging these two steps in one. 
While subspace methods reduce the order of the estimated VARX model via a model reduction step, in this paper regularization takes care of both stability and ``complexity'' (in terms of McMillan degree) at once, while estimating the VARX model itself. 

{\color{black}
Within this framework, our recent works \cite{PrandoCPCDC2014,Prando_SYSID15,PCCDC2015}  have attempted to merge the benefits of accounting for both stability/smoothness as well as complexity when building prior models. The main contributions of this work, w.r.t. the above referenced papers are:
(i) development, by means of MaxEnt arguments, of a new kernel encoding both complexity as well as smoothness and stability (the new kernel is parametrised differently w.r.t. previous conference publications and also the resulting algorithm is different); (ii) a new tailored Scaled Gradient Projection algorithm for marginal likelihood optimization (this had been used but not derived elsewhere) and (iii) an extensive simulation study comparing several state-of-the-art algorithms.\\
We shall now provide a more detailed description of these contributions as well as a brief discussion of the relevant literature.

 The first  main goal of this paper is to develop, by means of Maximum Entropy arguments,  a vector-valued kernel which accounts both for the stability of the system to be estimated and for its complexity, as measured by its McMillan degree.}
The prior distribution introduced here leads, as a special case,  to an Hankel nuclear norm penalty, an heuristic related to that  proposed in \cite{Fazel01} as a convex surrogate to the rank function. 
In the system identification literature the nuclear norm heuristic has also been applied in the context of subspace identification \cite{Hansson12,VerhaegenH2014,LiuV2009}, even in presence of incomplete datasets \cite{LiuHV13}, to control the order of the estimated model. PEM methods equipped with nuclear norm penalties on the Hankel matrix built with the Markov parameters have also been considered \cite{hjalmarsson2012identification,GrossmannJM09}. Refer to \cite{Prando_SYSID15} for a brief survey on the topic.\\
 However, direct use of nuclear norm (or atomic) penalties may lead to undesired behavior, as suggested and studied in \cite{PCCLDAuto2016}, due to the fact that nuclear norm is not able alone to guarantee stability and smoothness of the estimated impulse responses. To address this limitation,  \cite{Chiuso13}  already suggested the combination of the stability/smoothness penalty with the nuclear norm one; differently from the prior  presented in this paper, the formulation given in \cite{Chiuso13} did not allow to adopt marginal likelihood maximization to estimate the regularization parameters.\\
\noindent Exploting the structure of the prior distribution used in this paper we design an iterative procedure which alternatively updates the impulse response estimate and the hyper-parameters defining the prior. Our algorithm is related to 
iteratively reweighted methods used in compressed sensing and signal processing \cite{candes2008enhancing,chartrand2008iteratively,Daubechies_iterativelyreweighted,Mohan12jmlr,FornasierRW11} and so-called  \textit{Sparse Bayesian Learning} (SBL) \cite{WipfN10,Tipping01sparsebayesian}.\\
Our algorithm differs from the previous literature in that the regularization matrix takes on a very special structure, described by few hyper-parameters.
{\color{black} With  this special structure the weights update does not admit a closed-form solution and thus direct optimisation of the marginal likelihood needs to be performed.}\\ To this purpose, as a second main  contribution, this paper develops a  Scaled Gradient Projection method (SGP), inspired by the one introduced  in   \cite{BonettiniCPSIAM2014}, which is more efficient than  off-the-shelf optimization procedures implemented in MATLAB.\\
As a final contribution, the paper provides an extensive simulation study, where the proposed identification algorithm is compared with classical and state-of-the art identification methods, including PEM \cite{Ljung:99}, N4SID \cite{Vanov-book}, Stable Spline \cite{SS2011}, reweighted nuclear norm-based algorithms \cite{mohan2010reweighted} and regularized ``subspace'' methods \cite{VerhaegenH2014}.

{\color{black}While a clear-cut conclusion in terms of relative performance cannot be drawn at the moment,  it is fair to say that: (a) the new method developed in this paper outperforms the classical ``Stable-Spline'' \cite{SS2011}, especially when dealing with MIMO systems; (b) the new method outperforms  a reweighted Nuclear Norm algorithm in certain scenarios (e.g. a ``mildly-resonant'' fourth order system)  while performing comparably in others (e.g. randomly generated ``large''
MIMO systems).}

The paper is organized as follows. Section \ref{sec:PF} introduces the  problem and Section \ref{sec:Bayes} briefly frames system identification in the context of Bayesian estimation. In Section \ref{sec:ME} Maximum Entropy arguments are used to derive a family of  prior distributions. Section \ref{sec:Algorithm} illustrates our algorithm while Section \ref{sec:sgp} describes the adaptation of a Scaled Gradient Projection method, which is used to solve the marginal likelihood  optimization problem. An extensive experimental study will be conducted in Section \ref{sec:results}, while some concluding remarks will be drawn in Section \ref{sec:conclusion}.

\subsection*{Notation}
\noindent  In the following, $\mathbb{R},\mathbb{R}_+:=[0,\infty),\mathbb{Z}$ and $\mathbb{N}$ denote respectively the set of real, positive real, integers and natural numbers. $\mathbb{R}^n$ and $\mathbb{R}^{m\times n}$ will denote respectively the set of $n$-dimensional real vectors,  and $m \times n$ real matrices. The transpose of $A \in \R^{m\times n}$ will be denoted  $A^\top$. $0_n$, $0_{m\times n}$ and $I_{n}$ will denote respectively the zero  vector in $\R^n$,  the zero matrix in $\R^{m\times n}$ and the $n\times n$ identity matrix. The symbol $\otimes$ will denote the Kronecker product, $\mathcal{N}(\mu,\sigma)$ the  Gaussian distribution with mean $\mu$ and variance $\sigma$. Given $v\in\mathbb{R}^n$, $diag(v)$ will be a diagonal matrix of size $n\times n$ with the diagonal given by $v$. {\color{black} Given  matrices $V_i\in\mathbb{R}^{m_i\times n_i}$, $i=1,..,n$, $blkdiag(V_1,...,V_n)$ will denote the block-diagonal matrix of size $(m_1+...+m_n)\times (n_1+...+n_n)$ with the $V_i$'s as diagonal blocks.} $\mathbb{E}[\cdot]$ and $\mbox{Tr}\left\{\cdot\right\}$ will respectively denote  expectation and  trace.

\section{Problem Formulation}\label{sec:PF}
\noindent We consider the following linear, causal and time-invariant (LTI) Output-Error (OE) system:
\begin{equation}\label{equ:sys}
y(t) = H(q) u(t) + e(t)
\end{equation}
where $y(t) = [y_1(t),..,y_p(t)]^\top \in\mathbb{R}^p$ is the $p$-dimensional output signal, $u(t)= [u_1(t),..,u_m(t)]^\top\in\mathbb{R}^m$ is the $m$-dimensional input signal, $e(t)$ is additive noise and
\begin{equation}
H(q) = \sum_{k=1}^\infty h(k)q^{-k}
\end{equation}
is the system transfer function with $q^{-1}$ being the backward shift operator: $q^{-1}u(t)=u(t-1)$. For simplicity, we will assume the presence of a delay in $H(q)$, i.e. $h(0)=H(\infty)=0$. In addition, we assume $e(t)\sim \mathcal{N}(0_p,\Sigma)$, $\Sigma=diag(\sigma)$, $\sigma = [\sigma_1,...,\sigma_p]^\top$.

\noindent The objective is to estimate,  from a finite set of input-output data $\mathcal{D}_N=\left\{u(t),y(t);\ t=1,...,N\right\}$, the impulse response coefficients $\left\{h(k)\in\mathbb{R}^{p\times m};\ k=1,..., \infty\right\}$.\\
In the remaining of the paper, we shall consider $\left\{y(t);\ t\in\mathbb{Z}\right\}$ and $\left\{u(t);\ t\in\mathbb{Z}\right\}$ as jointly stationary zero-mean stochastic processes; furthermore, the input signal is assumed to be independent of the noise $\left\{e(t);\ t\in\mathbb{Z}\right\}$. The results of this paper can be easily extended to VARMAX/BJ type model structure, formulating the identification problem as estimation of the predictor model as done in \cite{SS2011}.

\section{Bayesian/regularization approach}\label{sec:Bayes}
\noindent In line with the recent developments in linear system identification, we tackle the problem outlined in Section \ref{sec:PF} by adopting a Bayesian approach. Namely, we consider $h$ as the realization of a stochastic process, embedding $h$ in an infinite-dimensional space. For simplicity,  consider  the Single-Input-Single-Output (SISO) case. A typical choice is to model $h$ as a zero-mean Gaussian process with covariance function $K_\eta:\mathbb{R}_+ \times \mathbb{R}_+ \rightarrow \mathbb{R}$,
\begin{equation}\label{equ:kernel}	
\mathbb{E}[h(t)h(s)] = K_{\eta}(t,s)
\end{equation}
where $K_\eta$ is parametrized via the hyper-parameter vector $\eta\in\Omega \subseteq \mathbb{R}^d$.
The covariance function $K_{\eta}(t,s)$, also called ``kernel'' in the machine learning literature, is appropriately designed in order to account for the desired properties of the impulse response to be estimated (e.g. stability, smoothness, etc.; see \cite{SS2011,ChenOL12,SurveyKBsysid}).\\
In this Bayesian framework the minimum variance estimate of $h$ conditional on the observations $\{y(t);\ t=1,...,N\}$, on the hyper-parameters $\eta$ and on the noise covariance $\Sigma$ is the conditional mean:
\begin{equation}\label{equ:min_var_est}
\widehat{h} := \mathbb{E}[h|Y,\eta,\sigma]
\end{equation}
where $Y\in\mathbb{R}^{Np}$ is the vector of output observations:
\begin{equation}\label{equ:y}
Y := \left[y_1(1) \ \cdots \ y_1(N)\ | \ \cdots\ | \ y_p(1) \ \cdots \ y_p(N)\right]^\top
\end{equation}
Assuming also that the noise $e$ is Gaussian and independent of $h$, $Y$ and $h$ will be jointly normal, so that for fixed $\eta$ and $\sigma$, $h$ conditioned on $Y$ is Gaussian. The estimator \eqref{equ:min_var_est} is then available in closed form; in particular, when $\eta$ and $\sigma$ are replaced with estimators $\hat \eta$ and $\hat\sigma$, \eqref{equ:min_var_est} is referred to as the Empirical Bayes estimate of $h$ \cite{Rasmussen}. 
Estimates of $\eta$ and $\sigma$ can be found  e.g. by  cross-validation or marginal likelihood maximization, i.e. 
\begin{equation}\label{equ:ML_max}
(\widehat{\eta},\widehat{\sigma}) := \arg \max_{\eta\in\Omega,\sigma\geq 0} p(Y\vert \eta,\sigma)
\end{equation}
where $p(Y\vert \eta,\sigma)$ denotes the likelihood of the observations $Y$ once the unknown $h$ has been integrated out, commonly called the \emph{marginal likelihood}. Under the Gaussian assumptions on $h$ and on the noise $e$ also the marginal likelihood $p(Y\vert \eta,\sigma)$  is a Gaussian distribution.

According to the Bayesian inference procedure outlined  above, the impulse response $h$ to be estimated lies in an infinite-dimensional space. However, thanks to the (exponentially) decaying profile of a stable impulse response, it is possible to estimate only a truncated version of $h$, i.e. to approximate $H(q)$ with the transfer function of a long Finite Impulse Response (FIR) model $H_{T}(q) = \sum_{k=1}^T h(k)q^{-k}$; in this way one avoids dealing with  infinite-dimensional objects. It should be stressed that the choice of the length $T$ does not correspond to a complexity selection step, since $T$ is simply taken large enough to capture the relevant dynamics of the unknown system. Henceforth, we will denote with $\mathbf{h}\in\mathbb{R}^{Tmp}$ the vector containing all the impulse response coefficients of $\{h(k);\ k=1,...,T\}$, appropriately stacked:
\begin{align}
\mathbf{h} &= [\mathbf{h}_{11}^\top\ \mathbf{h}_{12}^\top \ \cdots  \ \mathbf{h}_{1m}^\top \ \cdots\ \mathbf{h}_{p1}^\top \ \cdots \ \mathbf{h}_{pm}^\top]^\top \label{equ:h_vec}\\
\mathbf{h}_{ij}&=\left[h_{ij}(1)\ h_{ij}(2)\ \cdots\  h_{ij}(T)\right]^\top i= 1,..,p, \ j=1,..,m \nonumber
\end{align}
$h_{ij}(k)$ represents the $k$-th impulse response coefficient from input $j$ to output $i$. Under the Bayesian framework, $\mathbf{h}$ is a Gaussian random vector $\mathbf{h}\sim \mathcal{N}(0_{Tmp},\bar{K}_\eta)$, $\bar{K}_{\eta} \in\mathbb{R}^{Tmp \times Tmp}$. Exploiting the notation introduced in \eqref{equ:h_vec} and using the FIR approximation,  the convolution equation \eqref{equ:sys} can be reformulated as a linear model:
\begin{equation}\label{equ:vec_model}
Y = \Phi\mathbf{h} + E, \qquad \Phi\in\mathbb{R}^{Np \times Tmp},\ E\in\mathbb{R}^{Np}
\end{equation}
where the vector $E$ collects the noise samples, {\color{black} while the $\Phi= blkdiag(\phi,...,\phi)$ with $\phi\in\mathbb{R}^{N\times Tm}$ defined as:}
\begin{align}
\phi &= \left[\begin{array}{cccc}
\varphi_1(1) & \varphi_2(1) & \cdots & \varphi_m(1)\\
\varphi_1(2) & \varphi_2(2) & \cdots & \varphi_m(2)\\
\vdots & \vdots & \ddots & \vdots \\
\varphi_1(N) & \varphi_2(N) & \cdots & \varphi_m(N)\\
\end{array}\right] \nonumber\\
\varphi_i(j) &= \left[\begin{array}{cccc}
u_i(j-1) & u_i(j-2) & \cdots & u_i(j-T)
\end{array}\right]\nonumber\\
i &= 1,...,m, \quad j=1,...,N \label{equ:phi}
\end{align}
Since $\mathbf{h}$ and $E$ are modelled as  Gaussian and independent,  $Y$ and $\mathbf{h}$ are jointly Gaussian and  $\mathbf{h}$ conditionally on $Y$ is Gaussian, so that 
\eqref{equ:min_var_est} takes the form: 
\begin{equation}\label{equ:h_vec_hat}
\widehat{\mathbf{h}} := \mathbb{E}[\mathbf{h}\vert Y,\eta,\sigma] = \left[\Phi^\top \widetilde{\Sigma}^{-1}\Phi + \bar{K}_{\eta}^{-1}\right]^{-1}\Phi^\top \widetilde{\Sigma}^{-1}Y
\end{equation}
with $\widetilde{\Sigma} := \Sigma \otimes I_{N}$.
By recalling a known equivalence between Bayesian inference and regularization, the previous estimate can also be interpreted as the solution of the following Tikhonov-type regularization problem \cite{Wahba}:
\begin{equation}\label{equ:regul_probl}
\widehat{\mathbf{h}} = \arg\min_{\mathbf{h}\in\mathbb{R}^{Tmp}} (Y-\Phi\mathbf{h})^\top \widetilde{\Sigma}^{-1}(Y-\Phi\mathbf{h})+ J_\eta(\mathbf{h})
\end{equation}
with $J_\eta(\mathbf{h})=\mathbf{h}^\top \bar{K}_{\eta}^{-1} \mathbf{h}$.\\
The previous expression shows that the choice of the kernel $\bar{K}_\eta$ plays a crucial role for the success of the Bayesian inference procedure. Indeed, it shapes a regularization term $J_\eta(\mathbf{h})$ which penalizes impulse response coefficients corresponding to ``unlikely'' or ``undesired'' systems.
{\color{black} In Section 4 we will develop a new class of kernels which induces a penalty of the type:
\begin{equation}
J_{SH,\eta}(\mathbf{h}) = \mathbf{h}^\top \bar{K}_{S,\eta}^{-1} \mathbf{h} + \mathbf{h}^\top \bar{K}_{H,\eta}^{-1} \mathbf{h} 
\end{equation}
The first term in $J_{SH,\eta}(\mathbf{h})$ will account for the smoothness and stability of the impulse response to be estimated, while the second one will penalize high-complexity models. Estimation of the hyper-parameters $\eta$ and computation of the impulse response estimate $\widehat{\mathbf{h}}$ through an iterative algorithm will be discussed in Section \ref{sec:Algorithm}.}

\textcolor{black}{
\begin{remark}
The Bayesian inference scheme here illustrated has a well-known connection with the theory of Reproducing Kernel Hilbert Spaces (RKHS). Indeed, once a Gaussian prior for the impulse response $h$ is postulated with the covariance function defined in \eqref{equ:kernel}, the optimal estimate $\hat{h}$ can also be derived as the solution of a Tikhonov regularization problem and  will be an element of the RKHS $\mathbb{H}_{K_\eta}$  associated to the kernel $K_\eta$. If the true impulse response $h$ belongs to $\mathbb{H}_{K_\eta}$, then the so-called ``model bias'', accounting for the error between the true $h$ and its closest approximation in the hypothesis space, disappears (\cite{hastie_09}, Sec. 7.3). 
In particular, the RKHS associated to the so called stable spline kernel (adopted in the sequel) is very rich. 
For instance,  the impulse response of any BIBO stable finite dimensional linear system belongs to $\mathbb{H}_{K_\eta}$  for a suitable choice of $\eta$. In practice, $\eta$ is 
estimated by  \eqref{equ:ML_max}: this permits to tune model complexity, trading bias and variance\footnote{For the reason discussed above only ``estimation-bias'' will be present.}, in a  continuous manner.
\end{remark}
}

\section{Derivation of stable Hankel-type priors}\label{sec:ME}
\noindent In recent  contributions the standard smoothing spline kernels \cite{Wahba} have been adapted in order to represent covariances of exponentially decaying functions (\cite{SS2010}, \cite{SS2011}). For instance, considering SISO systems, the 1st order stable spline kernel (see \cite{SS2010} and \cite{ChenOL12} where it has been named Tuned-Correlated (TC) kernel) is defined as
\begin{equation}\label{equ:tc_kernel}
\left[\bar{K}_{S,\nu}\right]_{kl} = c\ \min\left\{\beta^k,\beta^l \right\}
\end{equation}
where  $\nu=[c,\beta]$, $c\geq 0$, $0\leq \beta \leq 1$ play the role of hyper-parameters. For a suitable choice of $\beta$, the impulse response of any BIBO linear system belongs a.s. to the RKHS associated to the kernel in \eqref{equ:tc_kernel}, see  \cite{SS2010}; thus, by adopting this kernel
the ``model bias'' 
 is  zero. Recently, \cite{ChenACCLP15} has shown that the kernel function from which \eqref{equ:tc_kernel} derives admits a Maximum Entropy interpretation. More specifically  it is the covariance function of a zero-mean Gaussian process defined over $\N_+$, which is the solution to the Maximum Entropy problem with constraints ($  k=1,...,\bar{k}\in \N_+)$
\begin{align}\label{equ:ss_constraint_minimal}
\mbox{Var} \left[h(k+1)-h(k)\right] &= c \left(\beta^{k}-\beta^{k+1} \right)
\end{align}
Exploiting a well-known result on Maximum Entropy distributions, see e.g. \cite[p. 409]{Cover}, the zero mean Gaussian prior with covariance \eqref{equ:tc_kernel} can also be derived by imposing the constraint\footnote{Note  that constraint  \eqref{equ:ss_constraint}  contains \eqref{equ:ss_constraint_minimal}.
}
\begin{align}\label{equ:ss_constraint}
\mathbb{E}\left[\mathbf{h}^\top \bar{K}_{S,\nu}^{-1}\mathbf{h} \right]&  = \bar c
\end{align}

When dealing with MIMO systems one needs to consider a block-kernel, with the $ji$-th block  $\bar{K}_{S,\nu}^{(ji)}\in \mathbb{R}^{T\times T}$ (the cross-covariance of  the impulse response from the $i$-th input and   the $j$-th output) defined e.g. as in \eqref{equ:tc_kernel}. In the recent literature, see e.g. \cite{ChiusoPAuto2012}, the cross terms (i.e.  $\bar{K}_{S,\nu}^{(ji)}$, $i\neq j$) have been set to zero. As we shall argue in a moment,  this assumption is often unreasonable.

In fact,  while smoothness is considered as a synonymous of ``simplicity'' in the machine learning literature, a system theoretic way to measure complexity is via the McMillan degree of $h$, i.e. the order $n$ of a minimal state space realization
\begin{equation}\label{equ:ss}
\begin{array}{rcl}
x(t+1) &=& A x(t) + Bu(t) \quad x(t) \in \R^n \\
y(t) & = & Cx(t) + e(t) 
\end{array}
\end{equation}
The impulse response coefficients   of \eqref{equ:ss}  are given by $h(k) = CA^{k-1}B \in \R^{p\times m}$,  a relation which couples  the  impulse responses $h_{ij}(k)$   as  $i$ and $j$ vary. This calls for prior distributions on $\mathbf{h}$ which encode this coupling. To this end, we first introduce the block Hankel matrix $\mathcal{H}(\mathbf{h})\in\mathbb{R}^{pr\times mc}$ given by:
\begin{equation}\label{equ:hankel}
\mathcal{H}(\mathbf{h})=\left[\begin{array}{ccccc} h(1) & h(2) & h(3) & \cdots & h(c)\\ h(2) & h(3) & h(4) & \cdots & h(c+1)\\ \vdots & \vdots & \vdots & \ddots & \vdots \\ h(r) & h(r+1) & \cdots & \cdots & h(r+c-1)\end{array}\right]
\end{equation}
A classical result from realization theory (see \cite{TetherTAC1970}  for  details) states that the rank of the block Hankel matrix $\mathcal{H}(\mathbf{h})$ equals the McMillan degree  of the system, if $r$ and $c$ are large enough. In this work $r$ and $c$ are chosen so that $r+c-1=T$ and the matrix $\mathcal{H}(\mathbf{h})$ is as close as possible to a square matrix.

From now on, to the purpose of normalization,  we shall consider a weighted version  $\widetilde{\mathcal{H}}(\mathbf{h})$ of  $\mathcal{H}(\mathbf{h})$:  
\begin{equation}\label{equ:weighted_hankel}
\widetilde{\mathcal{H}}(\mathbf{h}):= W_2^\top {\mathcal{H}}(\mathbf{h}) W_1^\top
\end{equation}
where $W_1$ and $W_2$ are chosen, see  \cite{Chiuso13}, so that the singular values of $\widetilde {\mathcal{H}}(\mathbf{h})$ are conditional canonical correlation coefficients between future outputs and near past inputs, given the future inputs and remote past inputs. 

\begin{remark}
For Gaussian processes, there is a one-to-one correspondence between the Canonical Correlation Analysis (CCA) and  mutual information. Indeed,  the mutual information between past ($y^-$) and future ($y^+$)  of a Gaussian  process $\{y(t)\}_{t\in \Z}$  is given by:
\begin{equation}\label{equ:cca}
\mathbb{I}(y^+;\,y^-) = -\frac{1}{2} \sum_{k=1}^n\log (1- \rho_k^2)
\end{equation}
where $\rho_k$ is the $k-th$ canonical correlation coefficient and $n$ is the McMillan degree of a minimal spectral factor of $y$. \\ This provides a clear interpretation of canonical correlations as well as of the impact of shrinking them in terms of mutual information. A similar interpretation holds for systems with inputs, relating conditional mutual information and  conditional canonical correlations, i.e. singular values of \eqref{equ:weighted_hankel} with the proper choice of $W_1$ and $W_2$.
%
\end{remark}



{\color{black}\subsection{Maximum Entropy Hankel priors} We shall now introduce a probability distribution $p(\mathbf{h})$ for $\mathbf{h}$, such that samples drawn from $p(\mathbf{h})$  have low rank (or close to low rank) Hankel matrices. To this purpose, we would like to favour  some of the singular values of $\mathcal{H}(\mathbf{h})$ to be (close to) zero: this can be achieved imposing constraints on the eigenvalues of the weighted matrix $\widetilde{\mathcal{H}}(\mathbf{h})\widetilde{\mathcal{H}}(\mathbf{h})^\top$. Let   $u_i(\mathbf{h})$ be the $i$-th singular vector of $\widetilde{\mathcal{H}}(\mathbf{h})\widetilde{\mathcal{H}}(\mathbf{h})^\top$.  To achieve our goal we shall constrain the (expected value) of the  corresponding singular value  $s_i^2(\mathbf{h}) $, i.e. 
\begin{align}
\E \left[s_i^2(\mathbf{h})\right]&=  \E \left[u_i(\mathbf{h})^\top  \widetilde {\mathcal{H}}(\mathbf{h}) \widetilde {\mathcal{H}}(\mathbf{h})^\top u_i(\mathbf{h})\right] \leq \omega_i \label{eq:bound:SV}
\end{align}
for $i=1,...,pr$. Here the expectation is taken w.r.t. $p(\mathbf{h})$, while the $\omega_i$'s play the role of hyper-parameters that will have to be estimated from the data\footnote{In fact, one shall not estimate directly the $\omega_i$'s, but rather the corresponding dual variables appearing in the MaxEnt distribution, i.e. the $\lambda_i$'s in  \eqref{equ:max_ent_prior}.}.}

In order to design $p(\mathbf{h})$,  we first assume that an estimate $\widehat{\mathbf{h}}$ of $\mathbf{h}$ is available. We shall see in Section \ref{sec:Algorithm} how this ``preliminary'' estimate of $\mathbf{h}$ arises as an intermediate step in an alternating minimization algorithm.\\
Thus, we consider the (weighted) estimated Hankel matrix $\widetilde{\mathcal{H}}(\widehat{\mathbf{h}})$ and its singular value decomposition
\begin{equation}
\widehat U \widehat S \widehat U^\top : = \widetilde{\mathcal{H}}(\widehat{\mathbf{h}}) \widetilde{\mathcal{H}}(\widehat{\mathbf{h}})^\top
\end{equation}
We can now reformulate the constraints \eqref{eq:bound:SV} as
\begin{equation}\label{con:SV}
 \E \left[\hat u_i^\top  \widetilde {\mathcal{H}}(\mathbf{h}) \widetilde {\mathcal{H}}(\mathbf{h})^\top \hat u_i \right]\leq \omega_i, \qquad i=1,...,pr
\end{equation}
where $\hat u_i$ denotes the $i$-th column of $\widehat U$. 
In this way we have fixed the  vectors $\hat u_i$, so that only $\widetilde {\mathcal{H}}(\mathbf{h})$ is random in \eqref{con:SV}.
%
%
Fixing the $\hat u_i$'s, which in general are not the (exact) singular vectors of the ``true'' Hankel matrix, introduces a perturbation on the constraint (and thus on the resulting   prior distribution). 
One way to make the constrains \eqref{con:SV} robust to such perturbations is to group estimated singular vectors into the so-called ``signal'' and ``noise'' subspaces\footnote{In fact, in this way perturbations ``within'' the signal and noise subspaces respectively have no effect.}.
%
To this purpose let us group the first $n$ singular vectors and partition $\widehat U$ and $\widehat S$ as follows:
\begin{equation}\label{group:SVD}
\widehat U =[\begin{array}{cc}  \widehat U_n & \widehat U_n^\perp\end{array}]  \qquad \widehat S= blkdiag(\widehat S_n,\ \widehat S_{n}^\perp)
\end{equation}
where $\widehat U_n\in\mathbb{R}^{pr \times n}$. Note that, while the  $\hat u_i$'s corresponding to small  singular values are likely to be very noisy, both the ``signal'' space spanned by the columns of  $\widehat U_n$, as well as that spanned by $\hat u_i$, $i=n+1,..,pr$ (i.e. the column space of $\widehat U_n^\perp$) are much less prone to noise;  this is easily derived from a perturbation analysis of the singular value decomposition which shows that the error in $\widehat U_n^\perp$  depends on the 
gap between 
the smallest singular value of $\widehat S_n$ and the largest one of $\widehat S_{n}^\perp$. 
In view of these considerations, we can relax the constraints \eqref{con:SV} by aggregating the ``signal'' components (i.e. the first $n$ singular vectors):
\begin{equation}\label{con:SV:signal}
\begin{array}{rl}
\E  \left[ {\rm Tr}\left\{ \widehat U_n^\top  \widetilde{\mathcal{H}}(\mathbf{h}) \widetilde{\mathcal{H}}(\mathbf{h})^\top\widehat U_n \right\} \right] 
& \leq  \sum_{i=1}^n \omega_i \end{array}
\end{equation}
where well known properties of the trace operator have been used.
Similarly, we group the constraints on the ``noise'' component (i.e. the last $pr-n$ singular vectors)
\begin{equation}\label{con:SV:noise}
\E  \left[ {\rm Tr}\left\{ \left(\widehat U^\perp_n\right)^\top  \widetilde{\mathcal{H}}(\mathbf{h})\widetilde{\mathcal{H}}(\mathbf{h})^\top\widehat U^\perp_n\right\} \right]  \leq \sum_{i=n+1}^{pr} \omega_i
\end{equation}
Exploiting a well known result \cite[p. 409]{Cover}, we can build the Maximum Entropy distribution subject to the constraints \eqref{con:SV:signal} and \eqref{con:SV:noise}:

\scalebox{.88}{\parbox{\columnwidth}{%
 \begin{align}
p_{\zeta}(\mathbf{h})& \propto  \exp\left( -\lambda_1 {\rm Tr}\left\{ \hat U_n^\top  \widetilde {\mathcal{H}}(\mathbf{h}) \widetilde {\mathcal{H}}(\mathbf{h})^\top\widehat U_n \right\}\right)\cdot \nonumber\\
 & \qquad \exp\left( -\lambda_{2}  {\rm Tr}\left\{ \left(\widehat U^\perp_n\right)^\top  \widetilde {\mathcal{H}}(\mathbf{h}) \widetilde {\mathcal{H}}(\mathbf{h})^\top\widehat U^\perp_n \right\} \right) \nonumber \\
& \propto  \exp\left(- {\rm Tr}\left\{ \widehat U^\top  \widetilde {\mathcal{H}}(\mathbf{h}) \widetilde {\mathcal{H}}(\mathbf{h})^\top\widehat U \ blkdiag( \lambda_1 I_{n}, \lambda_2 I_{pr-n}) 
\right\}  \right) \nonumber \\
& \propto  \exp\left(- {\rm Tr}\left\{  \widetilde {\mathcal{H}}(\mathbf{h}) \widetilde {\mathcal{H}}(\mathbf{h})^\top \widehat Q(\zeta) \right\}  \right) \label{equ:max_ent_prior}
\end{align}
}}

where $\zeta:=[\lambda_1,\lambda_2,n]$, $\lambda_1\geq 0,\ \lambda_2\geq 0$, and 
\begin{align}
\widehat Q(\zeta) := &\widehat U\ blkdiag( \lambda_1 I_{n}, \lambda_2 I_{pr-n}) \ \widehat U^\top \label{equ:reg:matrix}
\end{align}
\begin{remark}
We would like to  stress that the quality of the relaxation introduced in constraints \eqref{con:SV:signal} and \eqref{con:SV:noise} depends on the relative magnitude of the Hankel singular values. Using the ``normalized'' Hankel matrix \eqref{equ:weighted_hankel}  plays an important role here since its singular values, being canonical correlations, are  in the interval $(0,1]$. 
 On the other hand, the aggregation of the singular values along the ``noise'' subspace resembles the role played by the regularization factor in Iterative Reweighted methods \cite{chartrand2008iteratively,WipfN10}.{\color{black} We refer to Appendix \ref{app:connection_irw} for a thorough discussion on the connection between these methods and our approach.}
\end{remark}
\begin{remark}
Notice that $\widehat{Q}(\zeta)$ in \eqref{equ:reg:matrix} is the sum of two orthogonal projections 
$\widehat{Q}(\zeta)=\lambda_1 \widehat{U}_n\widehat{U}_n^\top + \lambda_2 \widehat{U}_n^\perp \left(\widehat{U}_n^\perp\right)^\top $, respectively on what we called the ``signal subspace'' (that would coincide with the column space of $\widetilde {\mathcal{H}}(\mathbf{h})$ if $n$ was the true system order) and on the ``noise subspace''. This observation provides new insights on the design of the prior in \eqref{equ:max_ent_prior}: namely, by properly tuning the hyper-parameters $\zeta$, the prior is intended to be stronger along certain directions of the column space of $\widetilde{\mathcal{H}}(\mathbf{h})$ (referred  to as the ``noisy'' ones) and milder along what we call the ``signal'' directions.
\end{remark}
Since $\widetilde {\mathcal{H}}(\mathbf{h})$ is linear in $\mathbf{h}$,   ${\rm Tr}\left\{ \widetilde {\mathcal{H}}(\mathbf{h}) \widetilde {\mathcal{H}}(\mathbf{h})^\top \widehat Q(\zeta) \right\}$ is quadratic  in $\mathbf{h}$ and letting $\widehat{Q}(\zeta)=LL^\top$, it can be rewritten as: \begin{align}
\mbox{Tr}&\left\{\widetilde{\mathcal{H}}(\mathbf{h})\widetilde{\mathcal{H}}(\mathbf{h})^\top \widehat{Q}(\zeta)\right\}= \mbox{Tr}\left\{L^\top \widetilde{\mathcal{H}}(\mathbf{h})\widetilde{\mathcal{H}}(\mathbf{h})^\top L\right\}\label{equ:trace_penalty}\\
 &= \|\mbox{vec}(\widetilde{\mathcal{H}}(\mathbf{h})^\top L)\|_2^2\nonumber\\
 &= \|(L^\top W_2^\top\otimes W_1)\mbox{vec}(\mathcal{H}(\mathbf{h})^\top)\|_2^2\nonumber\\
 &= \mathbf{h}^\top P^\top (W_2\widehat{Q}(\zeta) W_2^\top\otimes W_1^\top W_1)P\mathbf{h} \label{equ:rank_quad_form}
\end{align}
where $P\in\mathbb{R}^{rpcm\times Tmp}$  is such that 
$\mbox{vec}\left(\mathcal{H}(\mathbf{h})^\top\right) = P\mathbf{h}$. Inserting \eqref{equ:rank_quad_form} in \eqref{equ:max_ent_prior} we obtain 
\begin{equation}\label{equ:quadratic_hankel_prior}
p_\zeta(\mathbf{h}) \propto \exp\left(-\mathbf{h}^\top P^\top (W_2\widehat{Q}(\zeta) W_2^\top\otimes W_1^\top W_1)P\mathbf{h} \right)
\end{equation}
i.e. for  given $\zeta$, $\mathbf{h}$ is a  zero-mean Gaussian vector:
\begin{equation}\label{equ:hankel_prior}
\mathbf{h}\sim \mathcal{N}(0_{Tmp}, \bar{K}_{H,\zeta}) 
\end{equation}
with
\begin{align}
\bar{K}_{H,\zeta} &= \left[P^\top (W_2\widehat{Q}(\zeta) W_2^\top\otimes W_1^\top W_1)P \right]^{-1} \label{equ:hankel_kernel}\\
\zeta&=[\lambda_1,\lambda_2,n] \label{equ:hankel_hyperp}
\end{align}
{\color{black} Using \eqref{equ:hankel_prior} as a prior distribution for $\mathbf{h}$, we can recast the problem of estimating $\mathbf{h}$ under the framework outlined in Section \ref{sec:Bayes}. In particular, complexity (in terms of McMillan degree) is controlled by properly choosing the hyper-parameters $\zeta$, which can be done by marginal likelihood maximization as further discussed in Section \ref{sec:Algorithm}.}


{\color{black}
\begin{remark} From the regularization point of view \eqref{equ:regul_probl},
the penalty induced by the kernel \eqref{equ:hankel_kernel} can also be derived through a variational bound. We refer the reader to \cite{PrandoCPCDC2014} for more details about this derivation.
\end{remark}
}

{\color{black}
We shall notice that, when $\zeta^* = \left[\lambda^*,\lambda^*,0\right]$, quantity \eqref{equ:trace_penalty} (from which kernel \eqref{equ:hankel_kernel} arises) reduces to
\begin{align}
\mbox{Tr}\left\{ \widetilde{{H}}(\mathbf{h}) \widetilde{\mathcal{H}}(\mathbf{h})^\top \widehat Q(\zeta^*) \right\} &= \mbox{Tr}\left\{ \widetilde{\mathcal{H}}(\mathbf{h}) \widetilde{\mathcal{H}}(\mathbf{h})^\top \lambda^*I_{rp} \right\}\nonumber\\
 &=\lambda^* \sum_i s_i^2(\mathbf{h})\label{equ:approx_nn_penalty}
\end{align}
where $s_i(\mathbf{h})$ are the singular values of $\widetilde{\mathcal{H}}(\mathbf{h})$. Thus, the   nuclear norm penalty on the (squared) Hankel matrix can be derived from kernel \eqref{equ:hankel_kernel} as a special case, i.e. for a special choice of the hyper-parameters. The use of nuclear norm regularization is not new in  system identification: a comparison with the literature can be found in Appendix \ref{app:connection_nn}.
}

{\color{black}\subsection{Maximum Entropy stable-Hankel priors}  
\noindent As thoroughly discussed in \cite{PCCLDAuto2016},  the kernel arising from the ``Hankel'' constraint alone would not necessarily lead to stable models. In fact  given an unstable system and its finite Hankel matrix ${\mathcal H}$,   it is always possible to design a stable system whose finite Hankel matrix (of the same size as ${\mathcal H}$) has the same singular values of ${\mathcal H}$. In addition, the Hankel prior does not include information on the correlation among the impulse response coefficients (see \cite{PCCLDAuto2016}). Thus, as a final step,
we shall consider the Maximum Entropy distribution  \cite[p. 409]{Cover}, under both stability \eqref{equ:ss_constraint} and low complexity
(\eqref{con:SV:signal} and \eqref{con:SV:noise}) constraints, thus obtaining 
\begin{align}\label{equ:stable_hankel_prior}
p_{\eta}(\mathbf{h}) &\propto \exp\left(-\lambda_0\mathbf{h}^\top \bar{K}_{S,\nu}^{-1}\mathbf{h} - \mathbf{h}^\top \bar{K}_{H,\zeta}^{-1}\mathbf{h} \right)\nonumber\\
&\propto \exp\left(-\mathbf{h}^\top \left(\lambda_0\bar{K}_{S,\nu}^{-1} + \bar{K}_{H,\zeta}^{-1}\right)\mathbf{h} \right)
\end{align}
where $\eta=\left[\nu,\lambda_0,\zeta\right]$, $\lambda_0\geq 0$, and $\bar{K}_{H,\zeta}$ is the kernel in \eqref{equ:hankel_kernel}. The use of a further hyper-parameter, $\lambda_0$, will become clear later on. From the distribution \eqref{equ:stable_hankel_prior} we can derive the kernel 
\begin{align}\label{equ:ss_hankel_kernel}
\bar{K}_{SH,\eta} &= \left(\lambda_0 \bar{K}_{S,\nu}^{-1} + \bar{K}_{H,\zeta}^{-1}\right)^{-1}\\
&= \left[\lambda_0 \bar{K}_{S,\nu}^{-1} + P^\top (W_2\widehat{Q}(\zeta) W_2^\top\otimes W_1^\top W_1)P \right]^{-1}\nonumber
\end{align}
with hyper-parameters
\begin{equation}\label{equ:final_hyperp}
\eta= \left[\nu,\lambda_0,\zeta\right]
\end{equation}
and $\zeta$ as defined in \eqref{equ:hankel_hyperp}.
}

{\color{black} 
\section{Identification Algorithm} \label{sec:Algorithm}
This section describes the iterative algorithm to estimate the impulse response  $\mathbf{h}$ when the prior   is chosen as in \eqref{equ:stable_hankel_prior}. The algorithm alternates between the estimation of $\widehat{\mathbf{h}}$ (see \eqref{equ:h_vec_hat}) for fixed hyper-parameters and marginal likelihood optimization (see \eqref{equ:ML_max}).\\
The  procedure is summarized in Algorithm \ref{alg:ident}.
For ease of notation we have defined $\lambda:=\left[\lambda_0,\lambda_1,\lambda_2\right]$. Hence, the hyper-parameters vector $\eta$ in \eqref{equ:final_hyperp} can be rewritten as
$$\eta = \left[\nu,\lambda_0,\zeta\right] = \left[\nu,\lambda_0,\lambda_1,\lambda_2,n\right]= \left[\nu,\lambda,n\right]$$
Furthermore, $\widehat{\mathbf{h}}^{(k)}$, $\hat{\eta}^{(k)}$, $\hat{\lambda}^{(k)}$ and $\hat n^{(k)}$ denote estimators  at the $k$-th iteration of the algorithm.\\

\begin{remark}\label{rem:ss_hyperp}
In Algorithm \ref{alg:ident}  the noise variance $\sigma$ is fixed e.g. to the sample variance of an estimated ARX or FIR model.  
Of course $\sigma$ could also be treated as a hyper-parameter, and  estimated with  the same procedure based on the marginal likelihood.
\end{remark}

\begin{remark}
Algorithm \ref{alg:ident} has strong connections with iterative reweighted algorithms, see 
Appendix \ref{app:connection_irw} for details. \end{remark}

\begin{algorithm}
\caption{Identification Algorithm}
\label{alg:ident}
\begin{algorithmic}[1]
\STATE Set the resolution $\epsilon >0$
\STATE Estimate $\hat{\sigma}$ as illustrated in Remark \ref{rem:ss_hyperp}.
\STATE $\hat{n}^{(0)} \gets 0$ 
\STATE $\widehat{U}_{\hat{n}^{(0)}}\equiv \widehat{U}_0 \gets 0_{rp\times rp}$
\STATE $\widehat{U}_{\hat{n}^{(0)}}^\perp\gets I_{rp}$
\STATE $\hat{\nu} \gets \arg \max_{\nu\in\Omega} p(Y\vert \nu,\left[1,0,0\right],\hat{n}^{(0)},\hat{\sigma})$ \label{alg:est_nu}
\STATE $\hat{\lambda}^{(0)} \gets \arg \max_{\lambda\in\mathbb{R}_+^3} p(Y\vert \hat{\nu},\lambda,\hat{n}^{(0)},\hat{\sigma})$ \label{alg:ml_max_1}
\STATE $k \gets 0$
\WHILE{$\hat{n}^{(k)}<pr$} \label{alg:while_init}
\STATE $\widehat{\mathbf{h}}^{(k)} \gets \E [\mathbf{h}|Y,\hat{\eta}^{(k)},\hat{\sigma}]$ (using \eqref{equ:h_vec_hat}) \label{alg:h_estimate}
\STATE Compute the SVD: $\widetilde{\mathcal{H}}(\widehat{\mathbf{h}}^{(k)})\widetilde{\mathcal{H}}(\widehat{\mathbf{h}}^{(k)})^\top=
\widehat{U}\widehat{S}\widehat{U}^\top$
\STATE$\hat{n}^{(k+1)}\gets\hat{n}^{(k)}$
\LongState{Determine $\widehat{U}_{\hat{n}^{(k+1)}}$ and $\widehat{U}_{\hat{n}^{(k+1)}}^\perp$ from $\widehat{U}$.} \label{alg:compute_U}
\STATE $\hat{\lambda}^{(k+1)} \gets \arg \max_{\lambda\in\mathbb{R}_+^3} p(Y\vert \hat{\nu},\lambda,\hat{n}^{(k+1)},\hat{\sigma})$ \label{alg:ml_max_k}
\IF{{\tiny $p(Y\vert \hat{\nu},\hat{\lambda}^{(k+1)},\hat{n}^{(k+1)},\hat{\sigma}) > (1+\epsilon) p(Y\vert \hat{\nu},\hat{\lambda}^{(k)},\hat{n}^{(k+1)},\hat{\sigma})$}\small} \label{alg:ml_cond1} 
\STATE $k \gets k+1$
\ELSE
\STATE$\hat{n}^{(k+1)}\gets\hat{n}^{(k)}+1$
\STATE Perform steps \ref{alg:compute_U} to \ref{alg:ml_max_k}.
\IF{{\tiny{$p(Y\vert \hat{\nu},\hat{\lambda}^{(k+1)},\hat{n}^{(k+1)},\hat{\sigma}) > (1+\epsilon) p(Y\vert \hat{\nu},\hat{\lambda}^{(k)},\hat{n}^{(k+1)},\hat{\sigma}) \!\!\!$\small}}} \label{alg:ml_cond2} 
\STATE $k \gets k+1$  \label{alg:increasek} 

\ELSE
\STATE \textbf{break} \label{alg:break}
\ENDIF
\ENDIF
\ENDWHILE \label{alg:while_end}
\STATE Return $\widehat{\mathbf{h}} \gets  \widehat{\mathbf{h}} ^{(k)}$ \label{alg:h_final}
\end{algorithmic}
\end{algorithm}

Notice that the marginal likelihood maximization performed in steps \ref{alg:ml_max_1} and \ref{alg:ml_max_k} of Algorithm \ref{alg:ident} boils down to the following optimization problem:
\begin{align}
\hat{\lambda}^{(k)} = \arg\min_{\lambda\in\mathbb{R}_+^3} &Y^\top \Lambda(\hat{\nu},\lambda,\hat{n}^{(k)},\hat{\sigma})^{-1}Y +\nonumber\\
&\log\vert\Lambda(\hat{\nu},\lambda,\hat{n}^{(k)},\hat{\sigma})\vert \label{equ:ml_max_opt_probl}
\end{align}
where
\begin{equation}
\Lambda(\eta,\sigma) := \widetilde{\Sigma} + \Phi\bar{K}_{SH,\eta}\Phi^\top
\end{equation}
Section \ref{sec:sgp} will illustrate a Scaled Gradient Projection (SGP) method appropriately designed to solve  \eqref{equ:ml_max_opt_probl}.
We shall now discuss issues related to initialisation and convergence of Algorithm \ref{alg:ident}.

\subsection{Algorithm Initialization}
In the derivation of kernel $\bar{K}_{SH,\eta}$ in Section \ref{sec:ME} it has been assumed that a preliminary estimate $\widehat{\mathbf{h}}$ was available. Therefore the iterative algorithm we outline in this section has to be provided with an initial estimate $\widehat{\mathbf{h}}^{(0)}$. Exploiting the structure of the kernel $\bar{K}_{SH,\eta}$ in \eqref{equ:ss_hankel_kernel}, two straightforward choices are possible:
\begin{enumerate}
\item Initialize using only the stable-spline kernel (as the one in \eqref{equ:tc_kernel}), i.e.:
\begin{align}\label{equ:h_ini_SS}
\widehat{\mathbf{h}}^{(0)}&= \left(\Phi^\top \widetilde{\Sigma}^{-1}\Phi + \bar{K}_{S,\hat{\nu}^{(0)}}^{-1}\right)^{-1}\Phi^\top \widetilde{\Sigma}^{-1} Y \nonumber\\
\hat{\eta}^{(0)} &= \left[\hat{\nu}^{(0)},\hat{\lambda}^{(0)},0\right], \quad \hat{\lambda}^{(0)}=\left[1,0,0\right]
\end{align}
where only the hyper-parameters $\hat{\nu}^{(0)}$ are estimated through marginal-likelihood maximization \eqref{equ:ML_max}.
\item Initialize using the stable-Hankel kernel with $\hat n=0$, so that no preliminary estimate is needed to initialize $\hat U_n$ (which is empty) and thus $\hat Q(\hat \zeta^{(0)}) = \hat \lambda_2^{(0)} I$:  
\begin{align}\label{equ:h_ini_H}
\widehat{\mathbf{h}}^{(0)}&= \left(\Phi^\top \widetilde{\Sigma}^{-1}\Phi + \bar{K}_{SH,\hat{\eta}^{(0)}}^{-1}\right)^{-1}\Phi^\top \widetilde{\Sigma}^{-1} Y \nonumber\\
\hat{\eta}^{(0)} &= \left[\hat\nu^{(0)},\hat{\lambda}^{(0)},0 \right], \quad \hat{\lambda}^{(0)}=\left[1,\hat{\lambda}_2^{(0)},\hat{\lambda}_2^{(0)}\right]
\end{align}
where $\hat\nu^{(0)}$ and  $\hat{\lambda}_2^{(0)}$ are estimated through marginal likelihood maximization \eqref{equ:ML_max}.
\end{enumerate}
The procedure we actually follow (illustrated in Algorithm \ref{alg:ident}) combines the two strategies above. Namely, the first approach is adopted to fix the hyper-parameters $\hat{\nu}$ defining the stable-spline kernel (line \ref{alg:est_nu}). These are then kept fixed for the whole procedure. We then follow the second strategy to estimate $\hat{\lambda}^{(0)}$ (line \ref{alg:ml_max_1}). Note that in line \ref{alg:ml_max_1} the hyper-parameters ${\nu}$ are fixed to  $\hat{\nu}$ and not estimated as in \eqref{equ:h_ini_H}. Analogously, $\hat{\lambda}_0^{(0)}$ is estimated by marginal-likelihood maximization and not set a-priori to $1$ as in \eqref{equ:h_ini_H}. Therefore, the estimate $\widehat{\mathbf{h}}^{(0)}$ computed at line \ref{alg:h_estimate} is derived by adopting the kernel $\bar{K}_{SH,\hat{\eta}^{(0)}}$ with $\hat{\eta}^{(0)} = [\hat{\nu},\hat{\lambda}^{(0)},0]$.\\
This sort of ``hybrid'' strategy has been chosen for two main reasons. \emph{First}, it allows to fix the hyper-parameters ${\nu}$ by solving a simplified optimization problem (w.r.t. solving a problem involving all the hyper-parameters $\eta$). Notice that this also provides the user with a certain freedom on the choice of the kernel $\bar{K}_{S,\nu}$: using other kernel structures (see e.g. \cite{ChiusoCPL2014}) additional properties (e.g. resonances, high-frequency components, etc.) of the impulse response can be accounted for.
\emph{Second}, it also allows to properly initialize the iterative procedure used to update the hyper-parameters $\lambda_0$ and $\zeta$ in \eqref{equ:final_hyperp}, until a stopping condition is met (see next section for a discussion about convergence  of  Algorithm \ref{alg:ident}). \\

\subsection{Convergence Analysis}\label{rem:alg_convergence}
Algorithm \ref{alg:ident} is guaranteed to stop in a finite number of steps, returning a final estimate $\widehat{\mathbf{h}}$. Indeed, at any iteration $k$ four possible scenarios may arise: 
\begin{enumerate}
\item Condition at line \ref{alg:ml_cond1} is  met  and $k$ is increased by one and the algorithm iterates.
\item Condition at line \ref{alg:ml_cond1} is  not met\footnote{This  certainly happens after a finite number of iterations for any positive resolution $\epsilon$ and fixed $\hat n$.}, so that  $\hat n$ is increased by one, and   condition  \ref{alg:ml_cond2} is not met, then  the algorithm terminates returning  $\widehat{\mathbf{h}}:= \widehat{\mathbf{h}}^{(k)}$.
\item Condition at line \ref{alg:ml_cond1} is  not met$\,^{4}$, so that  $\hat n$ is increased by one, while condition  \ref{alg:ml_cond2} is  met,   
 then  $k$ is increased by one and the algorithm iterates.
 \item $\hat n^{(k)} = pr$, then the algorithm terminates returning $\widehat{\mathbf{h}}:= \widehat{\mathbf{h}}^{(k)}$.
 \end{enumerate}
Conditions (1) and (3) may only be satisfied a finite number of times, thus the algorithm terminates in a finite number of steps.\\
We also stress that Algorithm \ref{alg:ident} is only an ascent algorithm w.r.t. the marginal likelihood without any guarantee of convergence to a local extrema. If $\hat{U}_{\hat{n}^{(k)}}$ was treated as a hyper-parameter and the marginal likelihood optimised over the Grassmann manifold, then convergence to a local maxima  could be proven.\footnote{We have tested this variant, which is considerably more computationally expensive than  Algorithm 1. Since no significant improvements 
have been observed, we only present the simpler version in this paper.}
Notice indeed that we adopt a tailored Scaled Gradient Projection algorithm to solve the marginal likelihood optimization problem at line \ref{alg:ml_max_k} (see Section \ref{sec:sgp}): every accumulation point of the iterates generated by this algorithm is guaranteed to be a  stationary point (\cite{BonettiniCPSIAM2014}, Theorem 1); furthermore, for the specific problem we are solving, the sequence of the iterates admits at least one limit point.

Once the algorithm has converged, $\hat{n}$ is the optimal dimension of the ``signal'' and ``noise'' subspaces of $\widetilde{\mathcal{H}}(\mathbf{h})$, respectively spanned by the columns of $\widehat{U}_{\hat{n}}$ and $\widehat{U}_{\hat{n}}^\perp$. Furthermore, the corresponding multipliers $\lambda_1$ and $\lambda_2$ in $\zeta$ are expected to tend, respectively, to $0$ (meaning that no penalty is given on the signal component) and to $\infty$ (that is, a very large penalty is assigned to the noise subspace); if $\hat\lambda_2=\infty$, $\hat{n}$ would actually be the McMillan degree of the estimated system.\\
In practice the estimated   hyper-parameter $\hat\lambda_2$  is finite and, similarly,  $\hat\lambda_1$ is strictly positive. As a result the McMillan degree of the estimated system is generically larger than, but possibly close to, $\hat{n}$. Therefore, estimation of the   integer parameter $n$ should not be interpreted as a hard decision on the complexity as instead happens for parametric model classes whose structure is estimated with AIC/BIC/Cross Validation. Therefore, we may say that Algorithm \ref{alg:ident} performs  a ``soft'' complexity selection, confirming that this ``Bayesian''  framework allows to describe model structures in a continuous manner; in fact, for any choice of $\hat n$, systems of different McMillan degrees are assigned non zero probability  by the prior.

}

\section{SGP for marginal likelihood optimization}\label{sec:sgp}
A crucial step in Algorithm \ref{alg:ident} is the marginal likelihood maximization (step \ref{alg:ml_max_k}) 
which is  computationally expensive, especially  when the number of inputs and outputs  is  large. To deal with this issue we have adapted the Scaled Gradient Projection method (SGP), proposed in \cite{BonettiniCPSIAM2014}, to solve
\begin{align}
& \min_{\lambda\in\Omega}   f(\lambda)\label{eqn:opt} \\
f(\lambda) & = Y^\top \Lambda(\hat{\nu},\lambda,\hat{n},\hat{\sigma})^{-1}Y + \log\vert\Lambda(\hat{\nu},\lambda,\hat{n},\hat{\sigma})\vert
 \label{equ:sgp_obj}
\end{align}
with $\Omega = \mathbb{R}_+^3$. The SGP is a first order method, in which the negative gradient direction is doubly scaled through a variable step size  $\alpha_k$ and a positive definite scaling matrix $D_k$, which are iteratively updated. A careful choice of these scalings, illustrated later on,  allows to speed up   the, theoretically linear,    convergence; the reader is referred to \cite{BonettiniCPSIAM2014} for details.  The main steps of the algorithm are as follows (see Algorithm \ref{alg:sgp} for details):
\begin{enumerate}
\item  Set the descent direction (scaled  negative gradient)
\begin{equation}\label{equ:candidate_descent_dir}
\widetilde{\Delta\lambda}^{(k)}:= -\alpha_k D_k \nabla f(\lambda^{(k)})
\end{equation}
\item Project  the candidate update  $\tilde{\lambda}=\lambda^{(k)} +\widetilde{\Delta\lambda}^{(k)}$ on the constraint set 
\begin{equation}\label{equ:projection}
\Pi_{\Omega,D_k^{-1}}(\tilde{\lambda}) = \arg \min_{x\in\Omega} (x - \tilde{\lambda})^\top D_k^{-1} (x - \tilde{\lambda})
\end{equation}
and define the  final descent direction:
\begin{equation}\label{equ:final_descent_dir}
\Delta\lambda^{(k)}:= \Pi_{\Omega,D_k^{-1}}(\tilde{\lambda})  - \lambda^{(k)}
\end{equation}
{Since $\Omega$ is the positive cone  the projection is merely a truncation to non-negative values of $\tilde{\lambda}$ (and is independent of the scaling $D_k$)}.
\item Update $\lambda$ along the direction $\Delta\lambda^{(k)}$  as follows:
\begin{equation}\label{equ:param_update}
\lambda^{(k+1)} :=  \lambda^{(k)} + \delta_k \Delta\lambda^{(k)}
\end{equation}
with the steplength $\delta_k$  computed through an Armijo backtracking loop.
\end{enumerate}


\begin{algorithm}
\caption{Scaled Gradient Projection (SGP) method}
\label{alg:sgp}
\begin{algorithmic}[1]
\STATE Choose the starting point $\lambda^{(0)}\in\Omega$.
\STATE Set the parameters $\upsilon$, $\gamma\in(0,1)$, $0<\alpha_{min}<\alpha_{max}$, $0<L_{min}<L_{max}$ and the positive integer $M$.
\FOR{$k=0,1,2...$}
\LongState{Choose $\alpha_k\in\left[\alpha_{min},\alpha_{max}\right]$ and the diagonal scaling matrix $D_k$ such that $L_{min}<\left[D_k\right]_{ii}<L_{max}$, $i=1,2,3$.}\label{alg:alpha}
  {\color{black}\STATE Set the candidate direction as in \eqref{equ:candidate_descent_dir}}
{\color{black}\STATE Compute the projection $\Pi_{\Omega,D_k^{-1}}(\tilde{\lambda}) $ using \eqref{equ:projection}} \label{alg:direction}
{\color{black}\STATE Compute the descent direction as in \eqref{equ:final_descent_dir} }
\STATE Set $\delta_k=1$.
\IF{\small $f(\lambda^{(k)}+\delta_k\Delta \lambda^{(k)})\leq f(\lambda^{(k)})+\upsilon \delta_k \nabla f(\lambda^{(k)})^\top \Delta \lambda^{(k)}$} \label{alg:armijo}
\STATE Go to step \ref{alg:lambda_update}.
\ELSE
\STATE Set $\delta_k = \gamma\delta_k$ and go to step \ref{alg:armijo}.
\ENDIF
\STATE Set $\lambda^{(k+1)}=\lambda^{(k)}+\delta_k\Delta \lambda^{(k)}$. \label{alg:lambda_update}
\ENDFOR
\end{algorithmic}
\end{algorithm}

In  step \ref{alg:alpha} of Algorithm \ref{alg:sgp}, the stepsize $\alpha_k$ is chosen by means of an alternation strategy based on the Barzilai-Borwein rules (as  in \cite{BonettiniCPSIAM2014}), which  aim at finding  $\alpha_k$ so that $\alpha_k D_k$ approximates the inverse Hessian matrix.

The choice of the scaling matrix $D_k$ strictly depends on both the objective function and the constraints of the optimization problem. In our implementation we followed the choices made in \cite{BonettiniCPSIAM2014}: $D_k$ is set to be diagonal and its update is based on the split gradient idea. Let us first define
\begin{equation}
f_0(\lambda) = Y^\top \Lambda(\lambda)^{-1}Y, \qquad 
f_1(\lambda)= \log\vert\Lambda(\lambda)\vert
\end{equation}
where we have used the simplified notation $\Lambda(\lambda) \equiv \Lambda(\hat{\nu},\lambda,\hat{n},\hat{\sigma})$. {\color{black} Moreover, we define
\begin{equation}
\bar{K}_{SH,\lambda} := \left[\lambda_0\Gamma_0 + \lambda_1 \Gamma_1 +\lambda_2 \Gamma_2\right]^{-1} 
\end{equation}
%
where $\hat{\nu}$ and $\hat{n}$ are fixed and}
\begin{align}
\Gamma_0 &= \bar{K}_{S,\hat{\nu}}^{-1}\\
\Gamma_1 &= P^\top \left(W_2 \widehat{U}_{\hat{n}} \widehat{U}_{\hat{n}}^\top W_2^\top\otimes W_1^\top W_1\right)P \\
\Gamma_2 &= P^\top \left(W_2 \widehat{U}_{\hat{n}}^\perp \left(\widehat{U}_{\hat{n}}^\perp\right)^\top W_2^\top\otimes W_1^\top W_1\right)P 
\end{align}
Now, indicating with $\nabla_i f(\lambda)$ the gradient of $f$ w.r.t. to $\lambda_i,\ i=0,1,2$, we have:
\begin{align}
\nabla_i f_0(\lambda) &= Y^\top \Lambda(\lambda)^{-1} \Phi \bar{K}_{SH,\lambda} \Gamma_i \bar{K}_{SH,\lambda}\Phi^\top \Lambda(\lambda)^{-1} Y\\
\nabla_i f_1(\lambda) &= - \mbox{Tr}\left[\Phi^\top \Lambda(\lambda)^{-1}\Phi \bar{K}_{SH,\lambda} \Gamma_i \bar{K}_{SH,\lambda}\right]
\end{align}
From the positive definiteness of $\Lambda(\lambda)$ and the positive semidefiniteness of $\Phi \bar{K}_{SH,\lambda} \Gamma_i \bar{K}_{SH,\lambda}\Phi^\top$, it follows that $\nabla_i f_0(\lambda) \geq 0, \ \forall \lambda\in \mathbb{R}^3$.
Furthermore, from Lemma II.1 in \cite{lasserre1995trace}, it follows that $\nabla_i f_1(\lambda) < 0, \ \forall \lambda\in \mathbb{R}^3$.
This shows how the gradient of the objective function \eqref{equ:sgp_obj} admits the following decomposition:
\begin{equation}\label{equ:grad:split}
\nabla f(\lambda) = \nabla f_0(\lambda) + \nabla f_1(\lambda) = B(\lambda) - V(\lambda)
\end{equation}
with $B(\lambda) = \nabla f_0(\lambda) \geq 0$ and $V(\lambda) = - \nabla  f_1(\lambda) > 0$ (here the inequalities have to be understood component wise).					
Using  the gradient splitting \eqref{equ:grad:split}, the Karush-Kuhn-Tucker optimality conditions for problem \eqref{eqn:opt}
$$
\lambda_i\nabla_i f(\lambda)  = 0, \qquad \lambda_i\geq 0
$$
can be written as the solution of a fixed point iteration (see eq. (4.8) in \cite{BonettiniCPSIAM2014}) which leads to the scaling matrix  $D_k$:
\begin{equation}
\left[D_k\right]_{ii} = \min \left(\max\left(L_{min},\frac{\lambda_i^{(k)}}{V_i(\lambda^{(k)})}\right), L_{max}\right)
\end{equation}
This choice of the scaling matrix has proven to be particularly effective on ill-posed or ill-conditioned inverse problems, when it is combined with an appropriate choice of the stepsize $\alpha_k$. 


Further details on the setting of the parameters involved in Algorithm \ref{alg:sgp} and on the adopted stopping criterion will be given in Section \ref{subsec:sgp_perf}.

\section{Simulation Results} \label{sec:results}
\subsection{Monte-Carlo Simulations} \label{subsec:monte-carlo}
The identification procedure outlined in Algorithm \ref{alg:ident} is now compared with off-the-shelf identification routines, as well as with recently proposed methods. The comparison is performed through some Monte-Carlo studies on three appropriately designed  scenarios. \\
The innovation process $e(t)$ is always   a zero-mean Gaussian white noise with standard deviation randomly chosen in order to guarantee that the SNR on each output channel is a uniform random variable in the interval $[1,4]$. For each scenario we test the identification procedures on three different data lengths, which can be roughly classified as   ``few/average/many''  data. Each Monte-Carlo study includes $N_{MC}=200$ runs. A brief illustration of the three scenarios follows.
\begin{description}
\item[S1)]We consider a fixed fourth order system with transfer function $H(z)  = C(zI-A)^{-1}B$ where 
\begin{equation}\label{equ:sys_s1}
\begin{array}{c}
A = \mbox{blockdiag}\left(\left[\begin{array}{cccc}0.8 &0.5 \\ -0.5  & 0.8\end{array} \right],\left[\begin{array}{cccc}0.2 &0.9 \\ -0.9  & 0.2\end{array} \right]\right) \\
B = \left[ 1\; 0\; 2 \; 0\right]^\top  \quad C = \left[\begin{array}{cccc}  1 & 1 & 1 & 1 \\ 0 & 0.1 & 0 & 0.1 \\ 20 & 0 & 2.5 & 0\end{array} \right]
\end{array}
\end{equation}
The input is generated, for each Monte Carlo run, as a low
pass filtered white Gaussian noise with normalized band $[0,\varrho]$ where  $\varrho$ is a uniform random variable in the interval $[0.8, 1]$. The identification of system \eqref{equ:sys_s1} using data generated by a band-limited input appears particularly challenging because the system is characterized by two high-frequency resonances. \\The three different data lengths that have been considered are: $N_{1,1}=200,\ N_{1,2}=500,\ N_{1,3}=1000$.
\item[S2)] For each Monte Carlo run $H(z)$ is randomly generated using the MATLAB function \verb!drmodel! with $5$ outputs and $5$ inputs while guaranteeing that all the poles of $H(z)$ are inside the disc of radius $.85$ of the complex plane. The system orders are randomly chosen from 1 to 10. The input $u(t)$ is zero-mean unit variance white Gaussian noise. The three different numbers of input-output data pairs that have been tested are: $N_{2,1}=350,\ N_{2,2}=500,\ N_{2,3}=1000$.
\item[S3)] The systems have been randomly generated similarly to scenario   S2, but with  10 inputs and 5 outputs. Moreover, the input $u(t)$ is a low-pass filtered Gaussian white noise with normalized band defined as in S1. The considered data lengths are: $N_{3,1}=600,\ N_{3,2}=800,\ N_{3,3}=1000$.
\end{description}

\subsection{\color{black}Compared identification algorithms}\label{sec:compared_methods}
The following algorithms have been tested:\footnote{Some methods appeal to an Oracle (Or) who knows the true system. Clearly these are not feasible in practice and are only reported for the sake of comparison.}
\begin{enumerate}
\item N4SID+Or: The subspace method, as implemented by the MATLAB routine \verb!n4sid!. Different model complexities are tested;  an Oracle chooses the  order which maximises the impulse response fit \eqref{FIT}.
\item N4SID(OE)+Or: As N4SID+Or but forcing the routine to return an Output-Error model.
\item N4SID: The MATLAB routine \verb!n4sid!, equipped with default model order selection.
\item N4SID(OE): Same as N4SID by forcing an OE structure.
\item PEM+Or:   PEM as implemented by the MATLAB routine \verb!pem!. Different model complexities are tested: an Oracle chooses the  order which maximises the impulse response fit \eqref{FIT}.\item PEM(OE)+Or: Same as PEM+Or but using  the routine \verb!oe!. 
For each of the tested complexities, the routine \verb!oe! has been initialized with the model returned by \verb!pem!.

\item PEM: The MATLAB routine \verb!pem!, equipped with the default  model order selection.
\item PEM(OE): The MATLAB routine \verb!oe!, initialized with the model returned by \verb!pem! (order as selected by the default choice in \verb!pem!).
\item N2SID: The identification routine proposed in \cite{VerhaegenH2014} {\color{black}and implemented through the code available from \verb!http://users.isy.liu.se/en/rt/hansson/!}. This routine returns a state-space model in innovation form. The estimation of Output-Error models through N2SID has not been tested, since the routine does not straightforwardly allow to force an  OE model structure.
\item SS: The estimator \eqref{equ:h_vec_hat} where $\bar{K}_{\eta}$ is chosen to be the kernel TC introduced in \cite{ChenOL12}. {\color{black} The estimator is computed through the MATLAB routine \verb!arxRegul! (imposing a FIR model structure). }
\item NN+CV: A FIR model of order $T$ estimated solving
\begin{equation}\label{equ:nn_est}
\widehat{\mathbf{h}} = \arg\min_{\mathbf{h}\in\mathbb{R}^{mpT}} \|Y-\Phi\mathbf{h}\|_2^2 +\lambda^* \|\mathcal{H}(\mathbf{h})\|_*
\end{equation}
The optimization problem is solved through a tailored ADMM algorithm (as in \cite{LiuHV13}), while $\lambda^*$ is determined through Cross-Validation. This procedure has also been tested by replacing $\mathcal{H}(\mathbf{h})$ in \eqref{equ:nn_est} with $\widetilde{\mathcal{H}}(\mathbf{h})$ (see \eqref{equ:weighted_hankel}).
\item RNN+CV: A FIR model of order $T$ estimated by iteratively solving
\begin{equation}\label{equ:rew_nn_est}
\widehat{\mathbf{h}} = \arg\min_{\mathbf{h}\in\mathbb{R}^{mpT}} \|Y-\Phi\mathbf{h}\|_2^2 +\lambda^* \|W_l\mathcal{H}(\mathbf{h})W_r\|_*
\end{equation}
The weight matrices $W_l$ and $W_r$ are updated at each iteration according to the procedure suggested in \cite{mohan2010reweighted}. $\lambda^*$ is selected through Cross-Validation. The case in which $\mathcal{H}(\mathbf{h})$ in \eqref{equ:rew_nn_est} is replaced with $\widetilde{\mathcal{H}}(\mathbf{h})$ has also been tested.
\item SH: The estimator returned by Algorithm \ref{alg:ident} with $\bar{K}_{S,\nu}$ specified through the TC kernel. {\color{black} \footnote{\color{black}The MATLAB code is available upon request to the authors.}}
\end{enumerate}
Some implementation details follow. For  SS, SH, NN+CV and RNN+CV, the length $T$ of the estimated impulse response $\widehat{\mathbf{h}}$ has been set to 80 for scenario S1, to 50 for S2 and S3.
\\The regularization parameter $\lambda$  in N2SID \cite{VerhaegenH2014} has been chosen within a set of 20 elements logarithmically spaced between $10^{-3}$ and $10^{-1}$ for S1 and 40 elements logarithmically spaced between $10^{-3}$ and $10^{5}$ for S2 and S3.
The endpoints of these grids have been selected so that the estimated value of $\lambda$ is inside the interval.

{\color{black} When observed that using cross-validation, the results were unreliable for the ``few'' data  scenarios $N_{i,1}$.} To optimize the performance, in scenarios S2 and S3 we have used two-thirds of the available data as training set and the remaining one third for the validation step. Instead, in scenario S1, the available data have been equally split into the training and the validation set. The regularization parameter $\lambda^*$ has been selected from the vector $\tilde{v}=\frac{v}{N_{train}}$, where $N_{train}$ is the length of the training dataset, while $v$ is a vector of 25 elements logarithmically spaced between $10^{2}$ and $10^{7}$ for S1, between $10^{3}$ and $10^{7}$ for S2 and S3.

\subsection{\color{black}Impulse Response Estimate}
{\color{black} To evaluate the estimators  described above, we first introduce the so-called coefficient of determination (COD) between time series $a$ and $b$:
\begin{equation}\label{equ:cod}
\mbox{cod}_{N_c}(a,b) = 100 \left(1- \sqrt{\frac{\sum_{k=1}^{N_c} (a(k) - b(k))^2 }{\sum_{k=1}^{N_c} (a(k) - \bar{a})^2 } } \right)
\end{equation}
where $\bar{a}= \frac{1}{N_c}{\textstyle \sum_{k=1}^{N_c}} a(k)$. The impulse response fit is measured using the average COD:
\begin{equation}\label{FIT}
\mathcal{F}_{N_c}(\widehat{\mathbf{h}}) = \frac{1}{pm}\sum_{i=1}^p \sum_{j=1}^m \mbox{cod}_{N_c} (h_{ij}, \hat{h}_{ij})
\end{equation}
where $h_{ij}$ and $\hat{h}_{ij}$ denote the true and  estimated impulse responses from input $j$ to output $i$. We set $N_c=1000$, letting $\hat{h}_{ij}(k)=0,\ k=T+1,...,1000$.
}

\begin{figure}
\centering
\subfigure[]{
\includegraphics[width=\columnwidth]{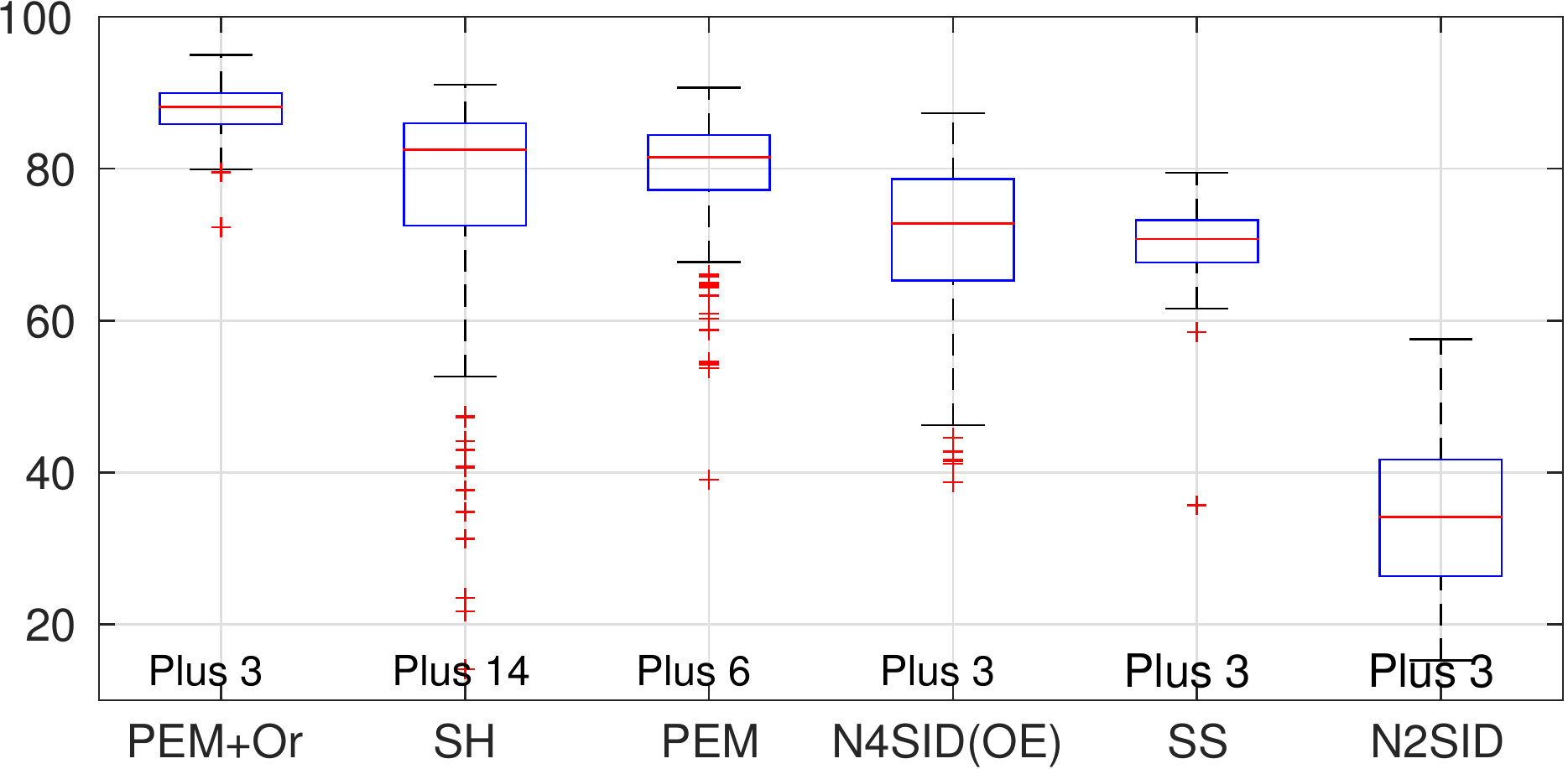}}
\subfigure[]{
\includegraphics[width=\columnwidth]{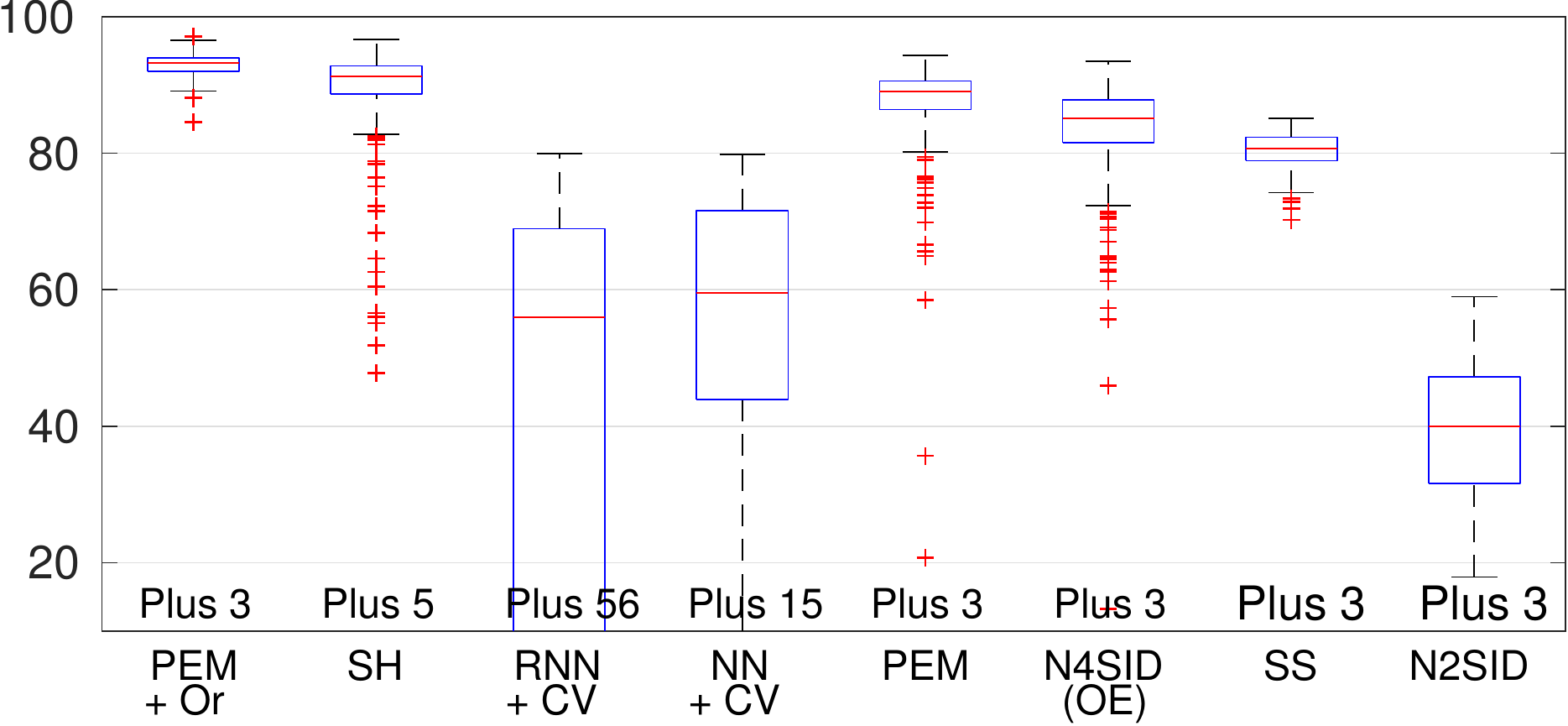}}
\subfigure[]{
\includegraphics[width=\columnwidth]{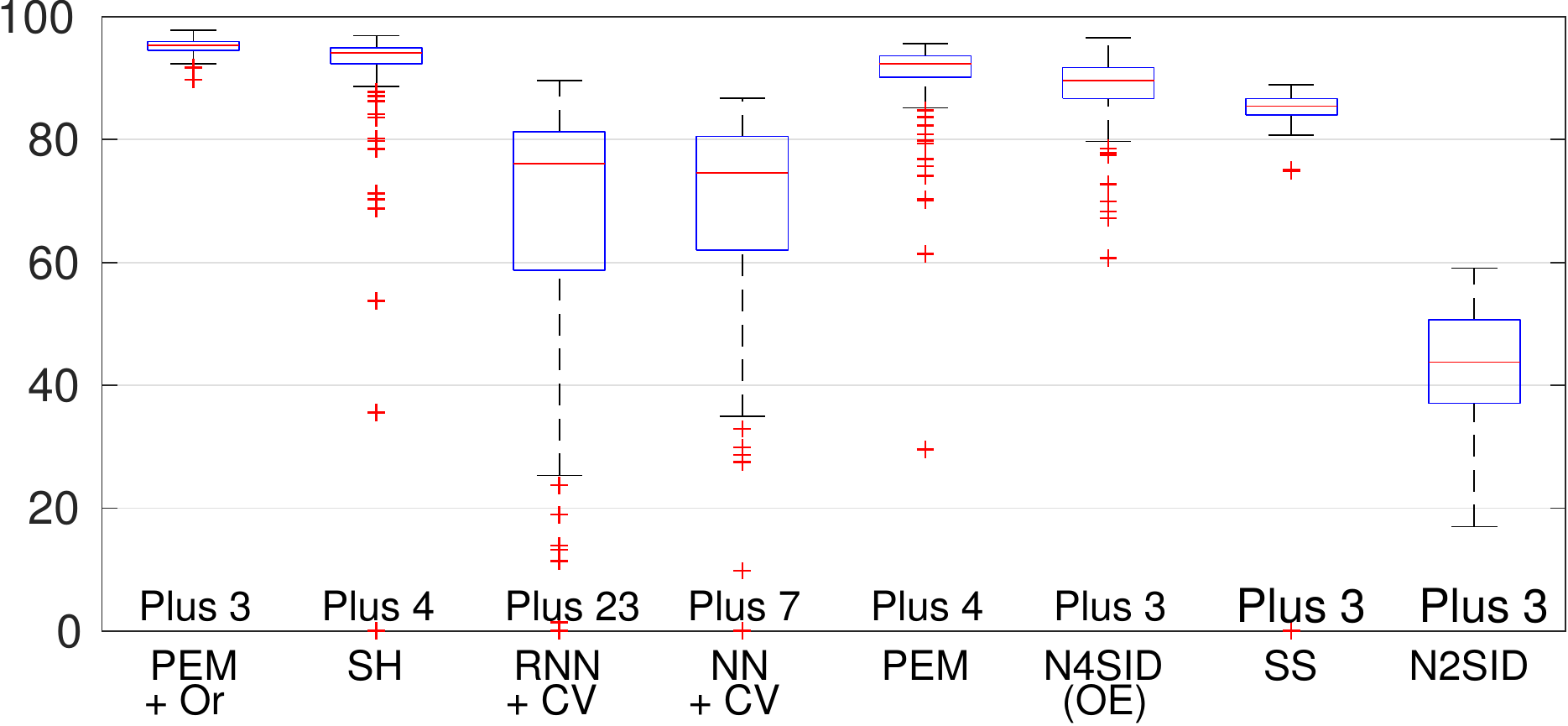}}
\caption{Scenario S1 - Fit \eqref{FIT} obtained with different data lengths: $N_{1,1}=200$ (a), $N_{1,2}=500$ (b) and $N_{1,3}=1000$ (c).}\label{fig:fit_s1}
\end{figure}

\begin{figure}
\centering
\subfigure[]{
\includegraphics[width=\columnwidth]{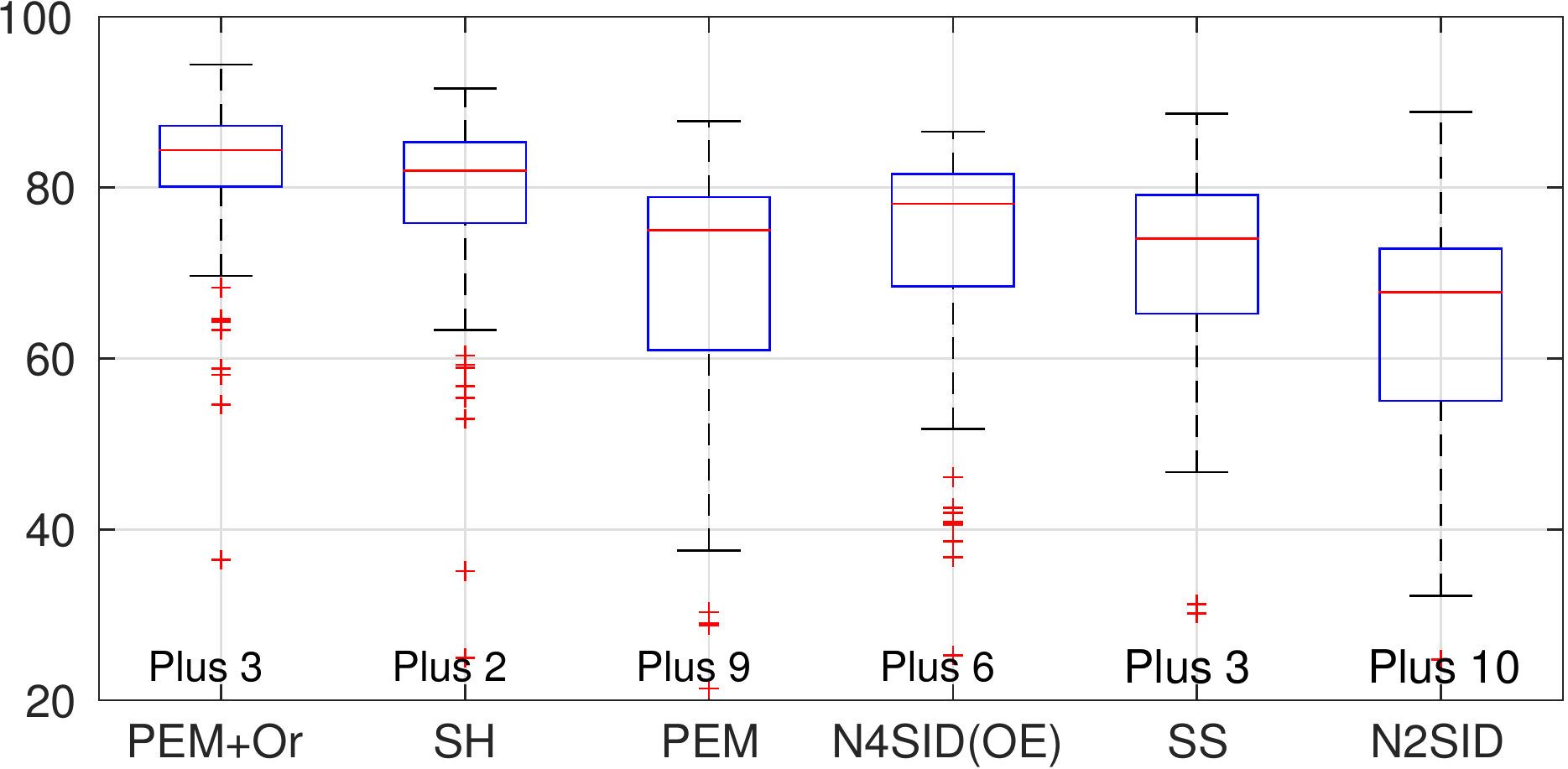}}
\subfigure[]{
\includegraphics[width=\columnwidth]{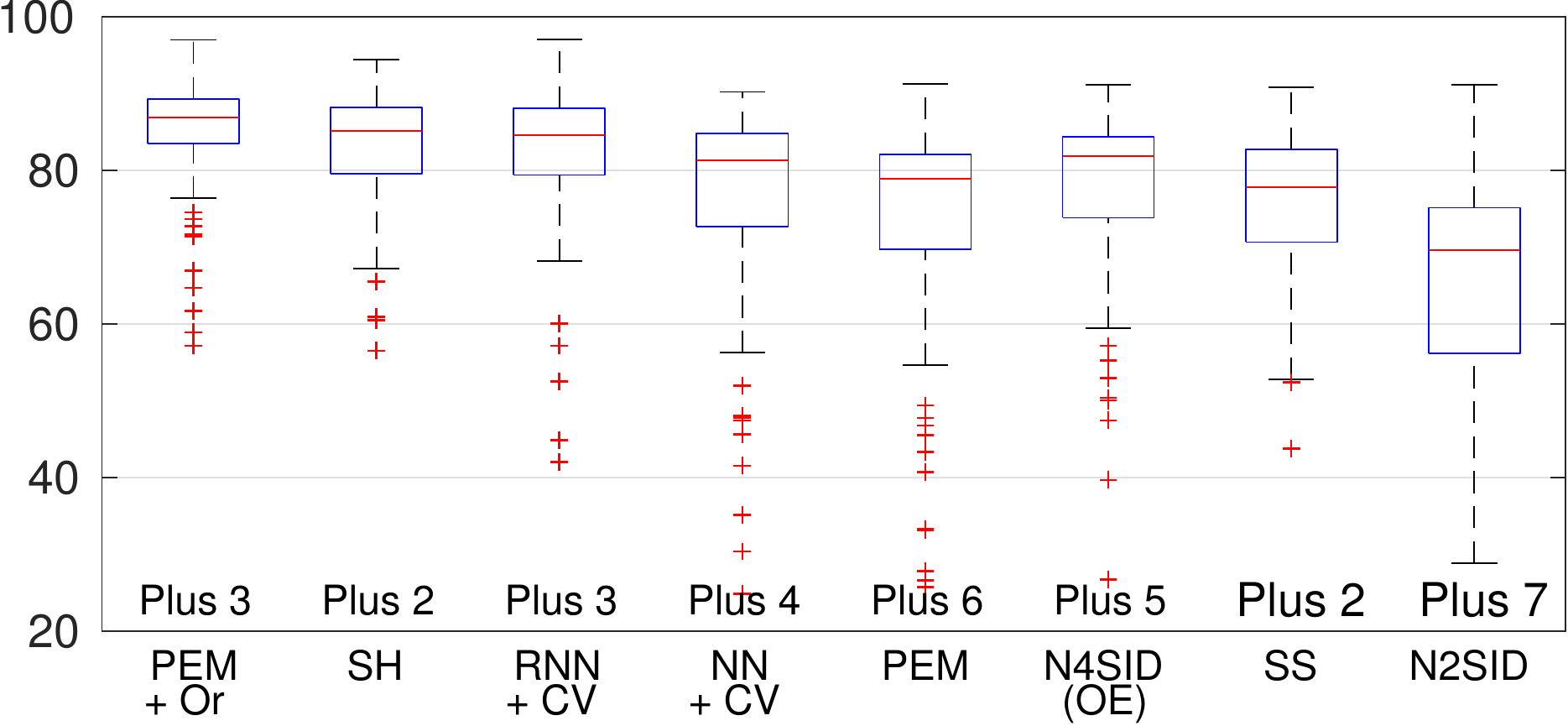}}
\subfigure[]{
\includegraphics[width=\columnwidth]{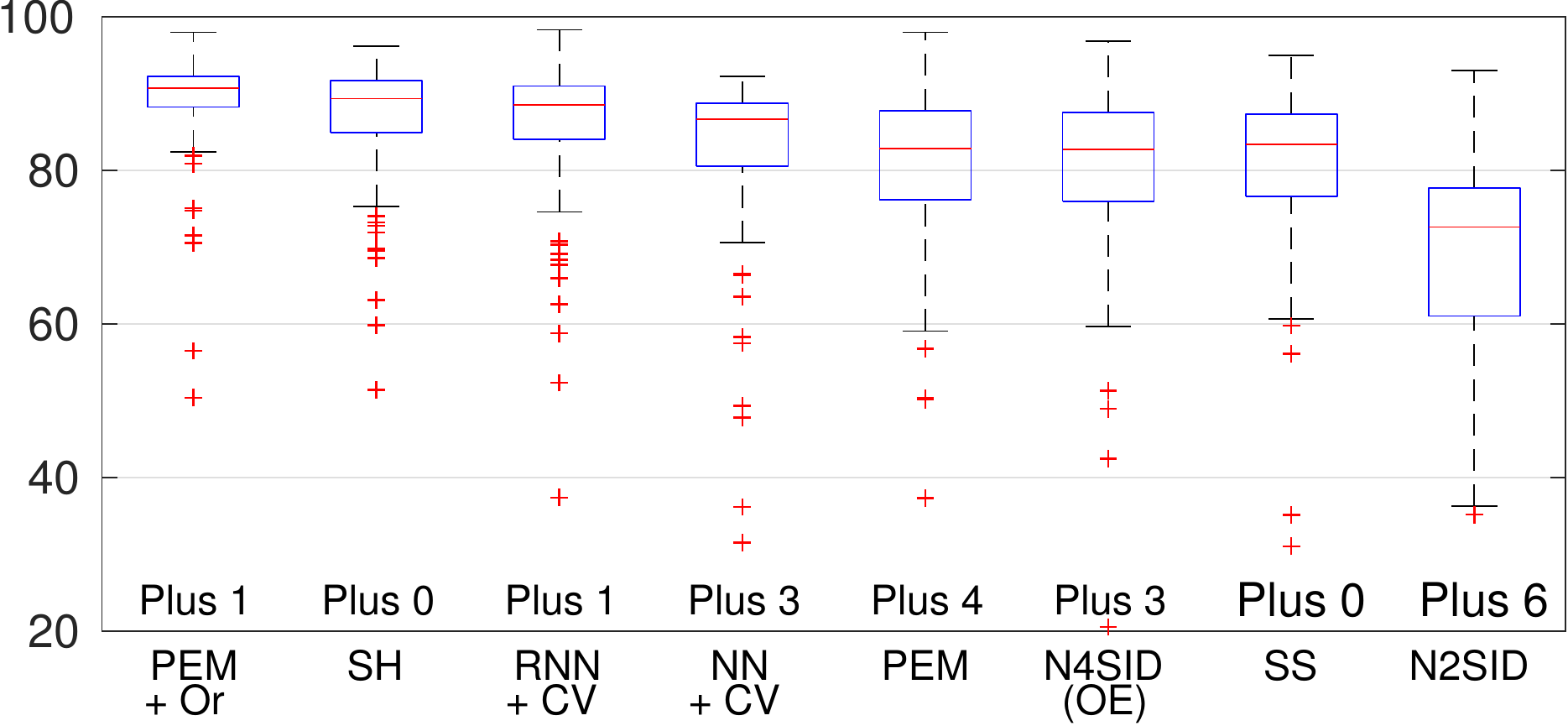}}
\caption{Scenario S2 - Fit \eqref{FIT} obtained with different data lengths: $N_{2,1}=350$ (a), $N_{2,2}=500$ (b) and $N_{2,3}=1000$ (c).}\label{fig:fit_s2}
\end{figure}

\begin{figure}
\centering
\subfigure[]{
\includegraphics[width=\columnwidth]{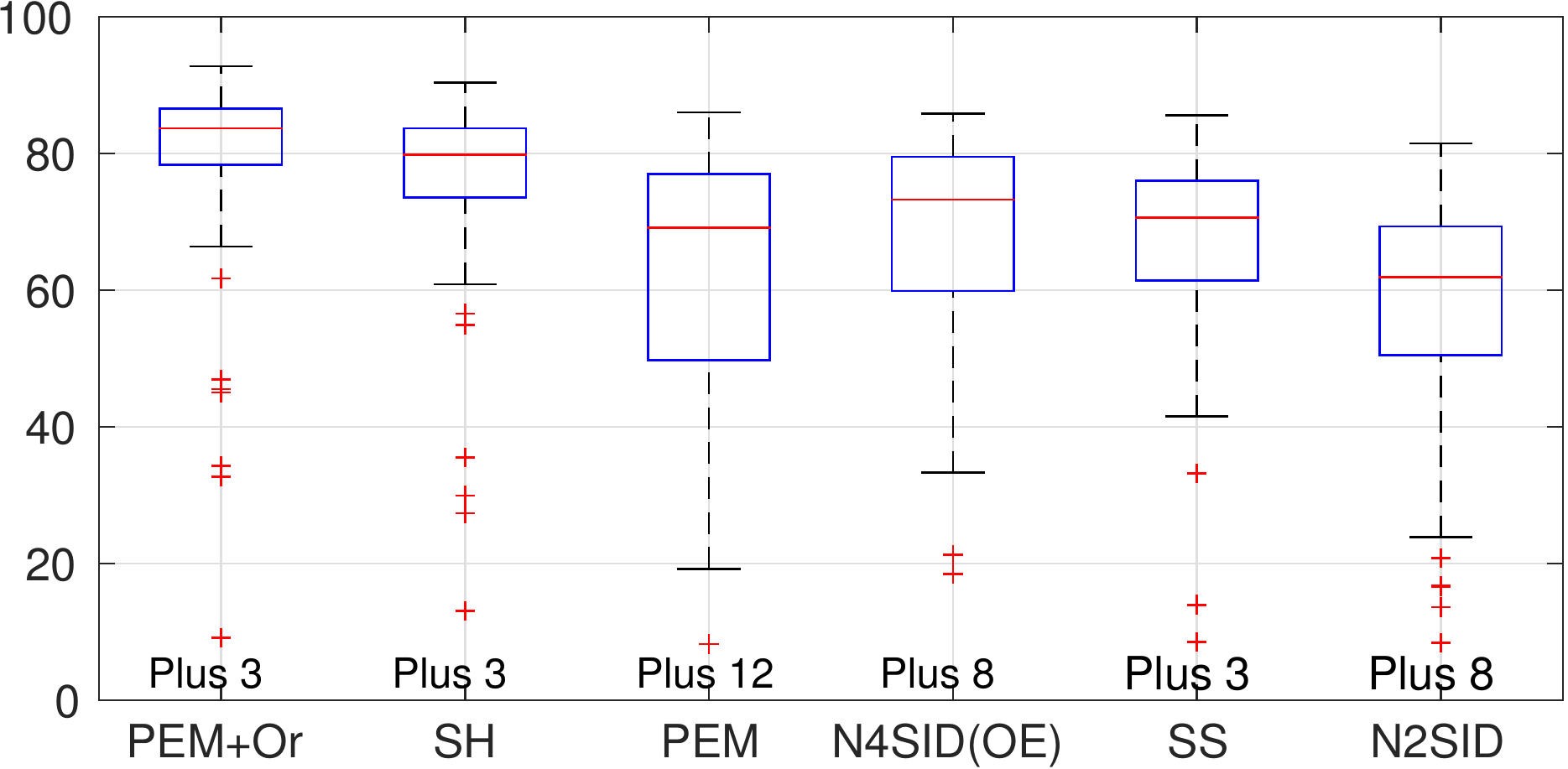}}
\subfigure[]{
\includegraphics[width=\columnwidth]{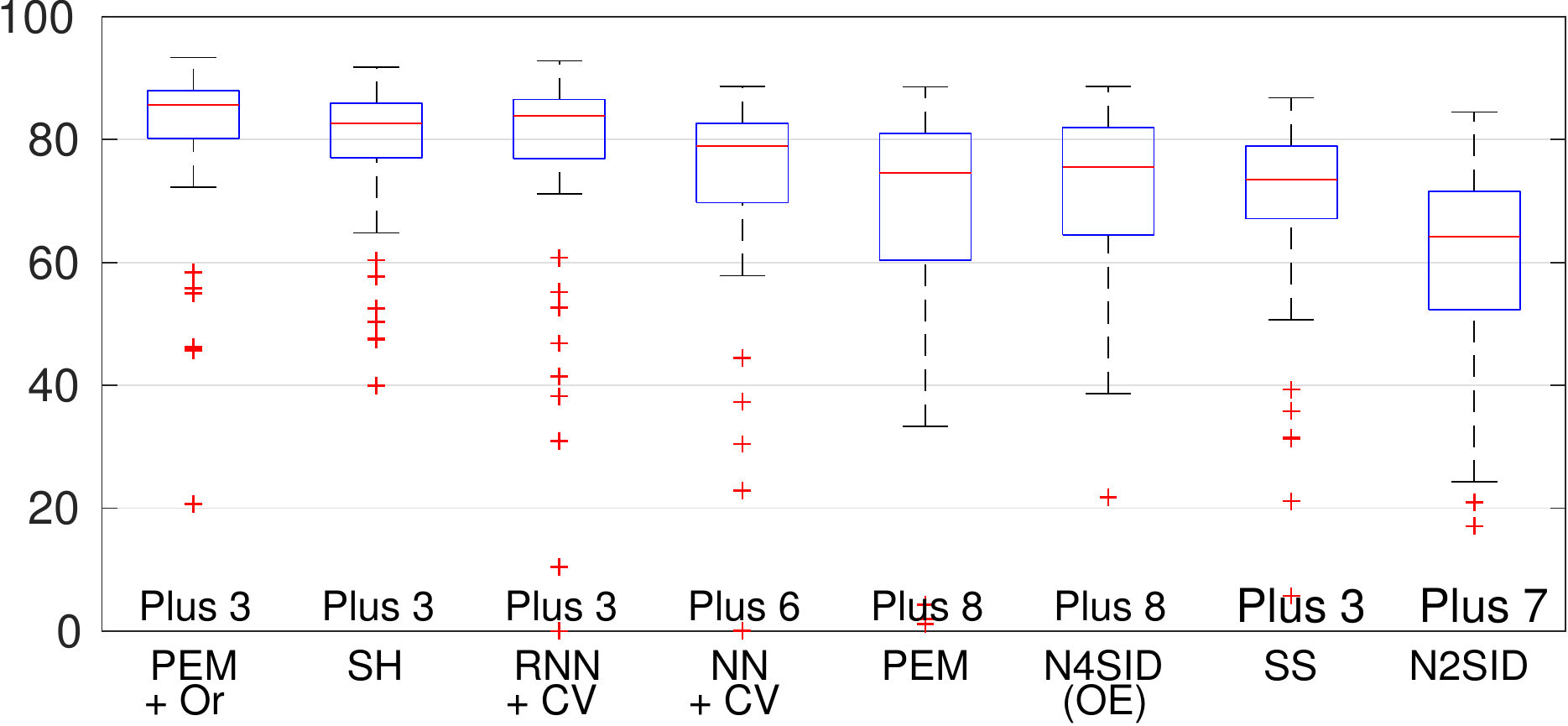}}
\subfigure[]{
\includegraphics[width=\columnwidth]{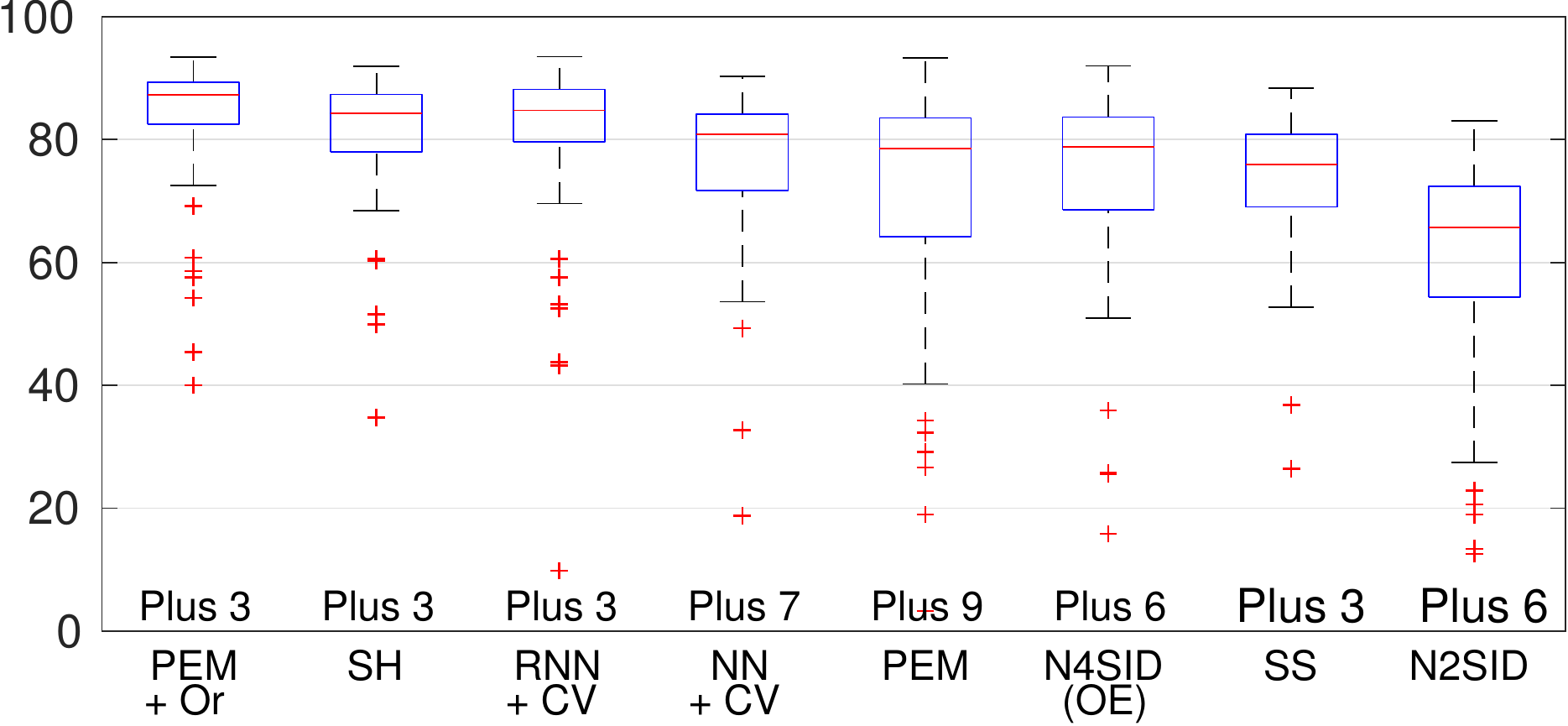}}
\caption{Scenario S3 - Fit \eqref{FIT} obtained with different data lengths: $N_{3,1}=600$ (a), $N_{3,2}=800$ (b) and $N_{3,3}=1000$ (c).}\label{fig:fit_s3}
\end{figure}

Figures \ref{fig:fit_s1}, \ref{fig:fit_s2} and \ref{fig:fit_s3} report the boxplots of \textcolor{black}{\eqref{FIT}} in the three scenarios by some of the identification techniques listed above. In particular, among the methods equipped with the oracle for  model complexity selection, only the results of PEM+Or are shown, since it gives the best performance. As far as  the subspace techniques are concerned, we only report N4SID(OE), because it generally performs slightly better than N4SID; analogously, only the results achieved by the routine PEM are illustrated, since the performance of PEM(OE) is worse.\\
SH and RNN+CV achieve, among the procedures which can be practically implemented,  the best performance in scenarios S2 and S3; instead, in scenario S1,  RNN+CV has severe difficulties. It is also interesting to observe that the reweighted procedure in \eqref{equ:rew_nn_est} (RNN+CV) improves the performance achieved by simple nuclear norm regularization (NN+CV) in all the scenarios except for S1. The results achieved imposing the nuclear norm penalty on the weighted Hankel matrix $\widetilde{\mathcal{H}}(\mathbf{h})$ are not reported since they are in general slightly worse than those achieved by NN+CV and RNN+CV.

\subsection{\color{black} Predictive performance}
{\color{black}We compare the predictive performance of the methods listed in Section \ref{sec:compared_methods} over a specifically designed scenario. Namely,  system \eqref{equ:sys_s1} has been simulated with a unit variance white Gaussian noise input, while its output was corrupted by additive white Gaussian noise with a variance chosen in order to have $SNR=2$. 200 estimation datasets consisting of $N=500$ data have been generated in this way. A set of validation data $\mathcal{D}_{N_{v}}^{v}=\{u^v(t), y^v(t); \ t=1,...,N_v \}$ was  used to evaluate the COD for each system output, i.e. $\mbox{cod}_{N_v}(y_i^v,\hat{y}_i),\ i=1,...,p$, (see definition in \eqref{equ:cod}) with $\hat{y}_i$ denoting the one-step ahead predictor for the $i$-th output channel.
Table \ref{tab:prediction} compares the median, the 5th and the 95th percentiles of $\mbox{cod}_{N_v}(y_i^v,\hat{y}_i)$ achieved by the considered identification methods.

}

\begin{table*}
\centering
\scriptsize
{\color{black}
\caption{Modified Scenario S1 - $\mbox{cod}_{N_v}(y_i^v,\hat{y}_i),\ N_v=500$ (see \eqref{equ:cod}): median, 5th and 95th percentiles over 200 Monte-Carlo runs. Estimators are computed using $500$ data (the best values among the realistic methods are highlighted in bold).}\label{tab:prediction}
\begin{tabular}{l|ccc|ccc|ccc}
\toprule
& \multicolumn{3}{c|}{$\mbox{cod}_{N_v}(y_1^v,\hat{y}_1)$} & \multicolumn{3}{c|}{$\mbox{cod}_{N_v}(y_2^v,\hat{y}_2)$}  & \multicolumn{3}{c}{$\mbox{cod}_{N_v}(y_3^v,\hat{y}_3)$} \\
& md & 5th pctl & 95th pctl & md & 5th pctl & 95th pctl & md & 5th pctl & 95th pctl \\      
\midrule
PEM+Or &92.54 & 87.69 & 95.94 & 92.76 & 88.77 & 96.14 & 92.74 & 88.06 & 95.86 \\ [-1.5ex]
SH & \textbf{91.48} & \textbf{86.85} & \textbf{95.03} & \textbf{91.55} & \textbf{86.60} & \textbf{95.31} & \textbf{91.46} & \textbf{86.80} & \textbf{94.83} \\ [-1.5ex]
RNN+CV &71.27 & 65.35 & 76.38 & 69.94 & 64.48 & 74.83 & 72.35 & 65.44 & 81.98 \\ [-1.5ex]
NN+CV &72.18 & 66.44 & 76.81 & 69.38 & 63.94 & 74.29 & 84.17 & 78.80 & 89.76 \\ [-1.5ex]
PEM &85.75 & 59.86 & 92.46 & 86.12 & 65.15 & 92.84 & 83.65 & 52.76 & 90.63 \\ [-1.5ex]
N4SID(OE) &82.42 & 70.05 & 89.71 & 81.85 & 66.69 & 90.22 & 88.36 & 80.80 & 92.39 \\ [-1.5ex]
SS &80.14 & 76.19 & 84.02 & 80.06 & 75.77 & 83.16 & 82.04 & 76.43 & 85.96 \\ [-1.5ex]
N2SID &34.78 & 11.85 & 51.04 & 26.59 & 7.57 & 43.34 & 58.95 & 49.26 & 65.79 \\ 
\bottomrule
\end{tabular}
}
\end{table*}


\subsection{\color{black}Analysis of estimated Hankel singular values}
\begin{figure}
\includegraphics[width=\columnwidth]{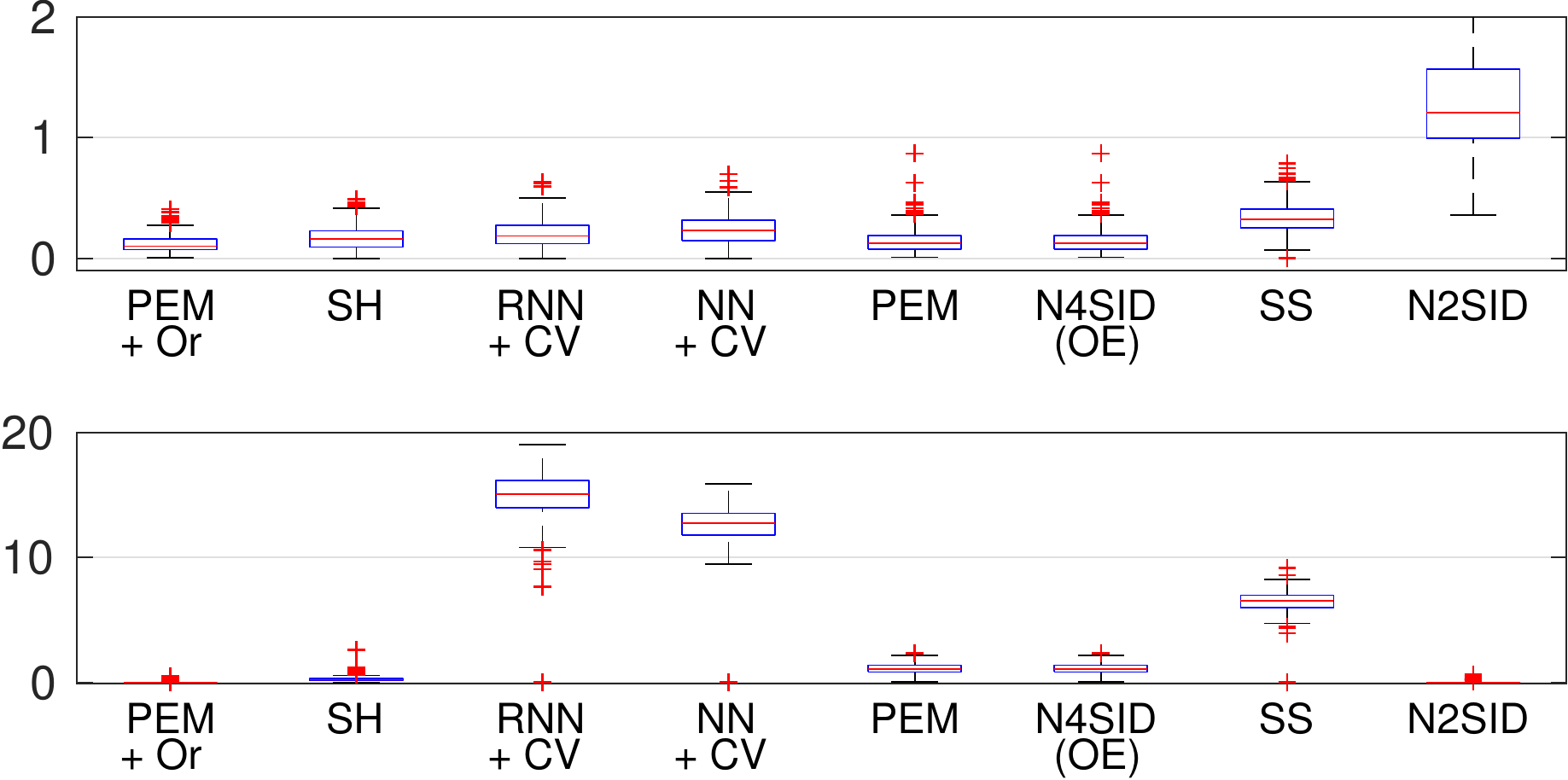}
\caption{Scenario S1 - \textit{Top}: Sum of absolute errors on the ``signal'' normalized Hankel singular values (see \eqref{equ:sum_err_signal_sv}). \textit{Bottom}: Sum of absolute errors on the ``noise'' normalized Hankel singular values (see \eqref{equ:sum_err_noise_sv}). Considered data length: $N_{1,2}=500$.}\label{fig:sv_s1}
\end{figure}

\begin{figure}
\includegraphics[width=\columnwidth]{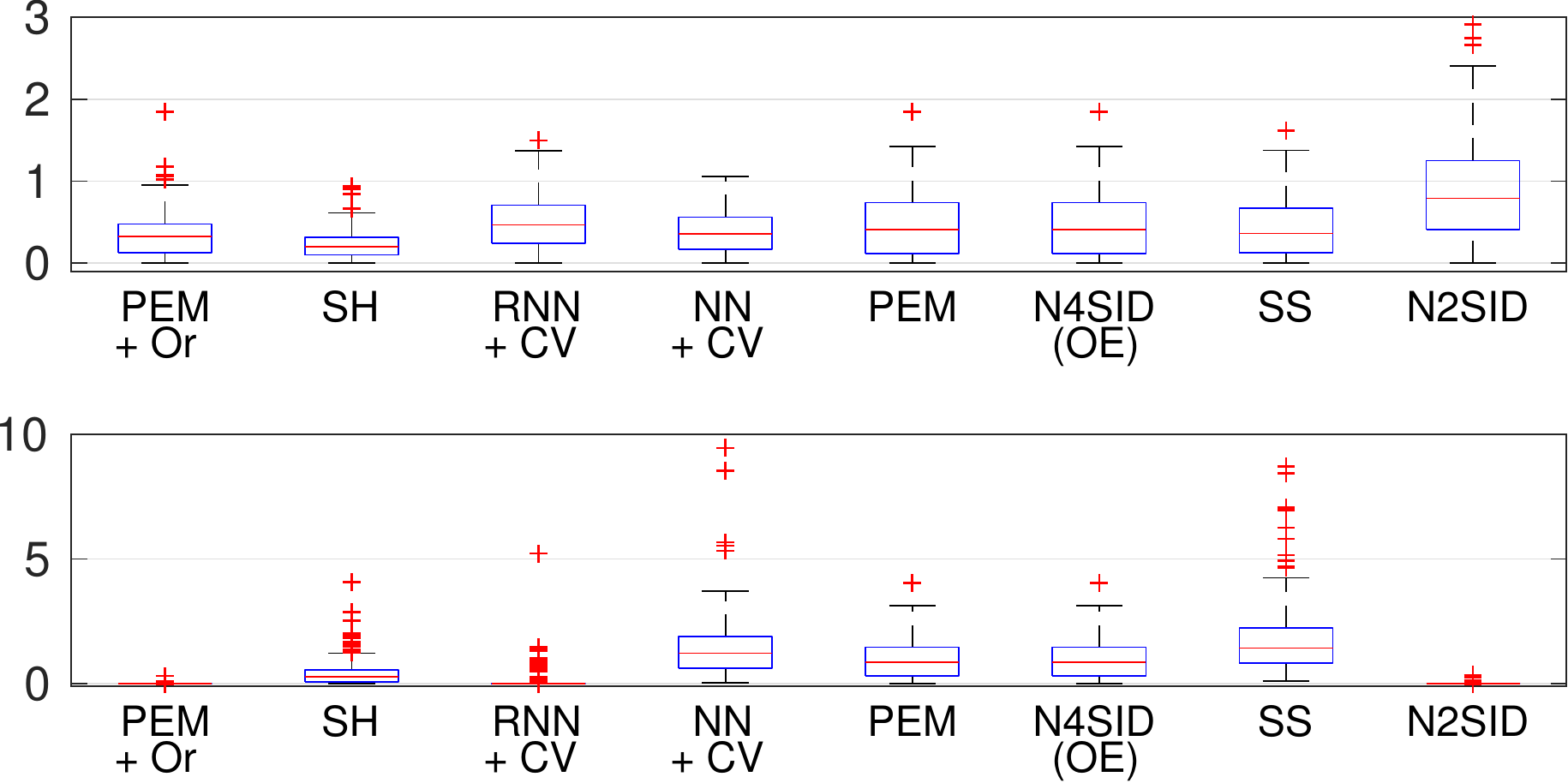}
\caption{Scenario S2 - \textit{Top}: Sum of absolute errors on the ``signal'' normalized Hankel singular values (see \eqref{equ:sum_err_signal_sv}). \textit{Bottom}: Sum of absolute errors on the ``noise'' normalized Hankel singular values (see \eqref{equ:sum_err_noise_sv}). Considered data length: $N_{2,2}=500$.}\label{fig:sv_s2}
\end{figure}

\begin{figure}
\includegraphics[width=\columnwidth]{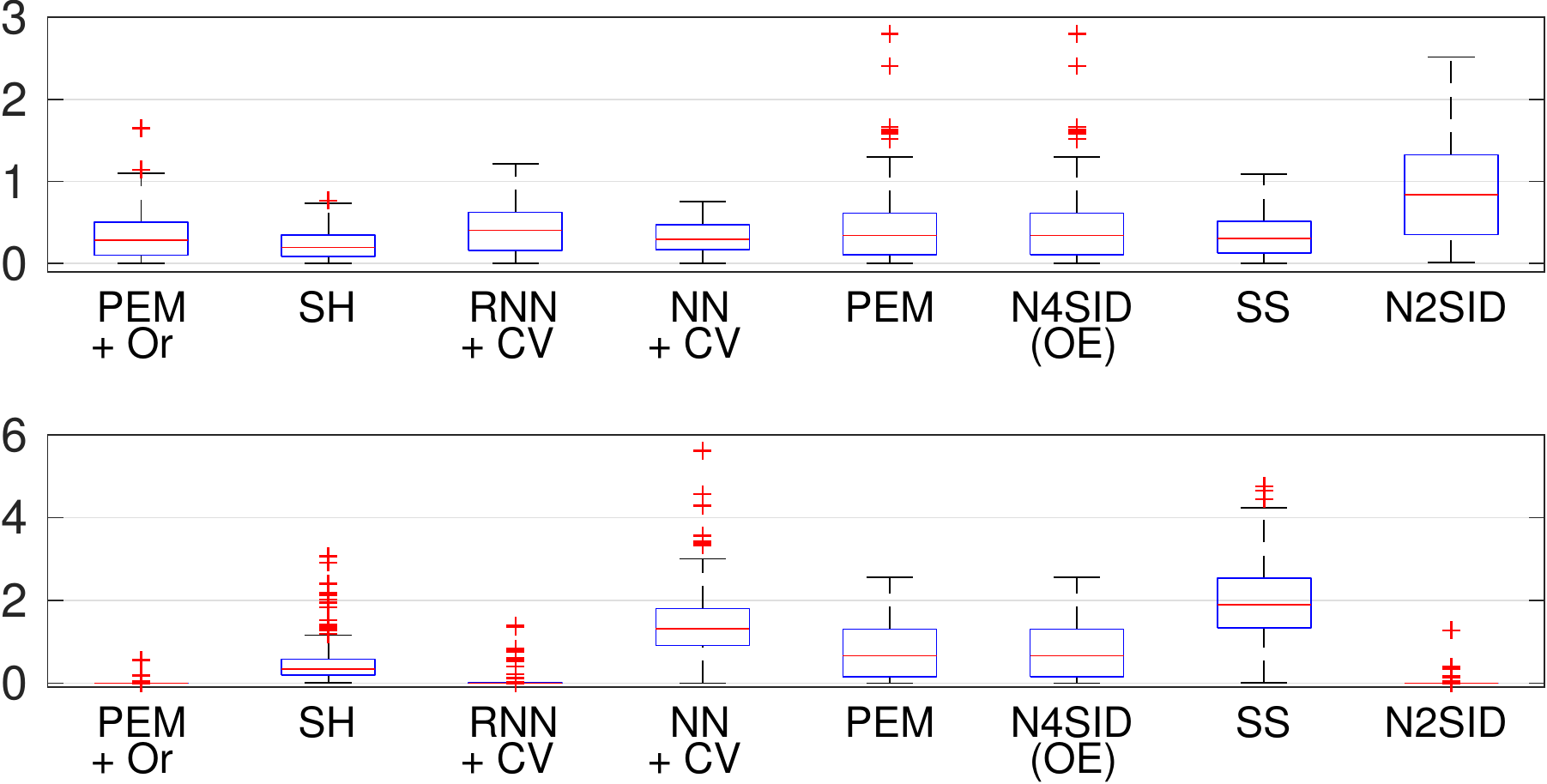}
\caption{Scenario S3 - \textit{Top}: Sum of absolute errors on the ``signal'' normalized Hankel singular values (see \eqref{equ:sum_err_signal_sv}). \textit{Bottom}: Sum of absolute errors on the ``noise'' normalized Hankel singular values (see \eqref{equ:sum_err_noise_sv}). Considered data length: $N_{3,2}=800$.}\label{fig:sv_s3}
\end{figure}

Figures \ref{fig:sv_s1}, \ref{fig:sv_s2} and \ref{fig:sv_s3} are concerned with the ability  in estimating the  Hankel singular values, which are grouped in the so called ``signal singular values'' (corresponding to the nonzero singular values of the true system) and ``noise singular values'' (corresponding to the zero singular vaues of the the true system). Indeed, the top plot in each figure shows the boxplots of the error on the ``signal singular values'': \begin{equation}\label{equ:sum_err_signal_sv}
\Delta_{signal}(\widehat{\mathbf{h}}):=\sum_{i=1}^{\bar{n}}|\tilde{s}_i(\mathbf{h}) - \tilde{s}_i(\mathbf{\widehat{h}})|
\end{equation}
where $\mathbf{h}$ is the true impulse response vector, $\widehat{\mathbf{h}}$ is the estimated one, 
 $\tilde{s}_i(\mathbf{h})$ is the $i$-th normalized Hankel singular value and  $\bar{n}$ here denotes the true system order.
 Similarly,  the bottom plot contains the boxplots of the error on the ``noise singular values'': 
 \begin{equation}\label{equ:sum_err_noise_sv}
\Delta_{noise}(\widehat{\mathbf{h}}): =\sum_{i=\bar{n}+1}^{pr}|\tilde{s}_i(\mathbf{h}) - \tilde{s}_i(\mathbf{\widehat{h}})|= \sum_{i=\bar{n}+1}^{pr}\tilde{s}_i(\mathbf{\widehat{h}}) \end{equation}
Figure \ref{fig:sv_s1} shows that the poor performance observed in Figure \ref{fig:fit_s1} for NN+CV and RNN+CV is determined by the failure in detecting the ``true'' system complexity (as proven by the large error in the estimation of the  ``noise'' singular values which can be interpreted as  overestimation of the system order). On the other hand, the unsatisfying performance of N2SID in Figure \ref{fig:fit_s1} is due to the under-estimation of the system complexity, which leads to a large bias in the estimation of the true Hankel singular values (top of Figure \ref{fig:sv_s1}) and to the correct detection of the ``noise'' subspace. Among the feasible methods, SH seems to correctly estimate the system complexity in most cases. \\
With regards to scenarios S2 and S3, the joint analysis of Figures \ref{fig:fit_s2}, \ref{fig:sv_s2} and \ref{fig:fit_s3}, \ref{fig:sv_s3} reveals how the good performance in terms of impulse response fit achieved by PEM+Or and RNN+CV are mainly due to the correct reconstruction of the ``noise'' subspace; indeed, the  performance of SH in terms of fit are slightly worse even if it better recovers the ``signal'' subspace. A deeper inspection has revealed that the system complexity is underestimated by PEM+Or, RNN+CV and N2SID, thus explaining the almost perfect reconstruction of the ``noise'' subspace and the bias which affects the estimates of the ``signal'' subspace. This observation suggests that the good performance observed for RNN+CV  in Figures \ref{fig:fit_s2} and \ref{fig:fit_s3} are favored by the nature of the systems in scenarios S2 and S3: indeed, underestimation of the system order does not have a detrimental effect in these scenarios where there are many ``small'' Hankel singular  values.\\
Comparing the performance of NN+CV and RNN+CV in Figures \ref{fig:sv_s2} and \ref{fig:sv_s3}, it is clear that the reweighted procedure significantly increases the degree of sparsity in the estimated Hankel singular values.

\subsection{Computational time}\label{subsec:sgp_perf}
{\color{black} A comparison of the methods listed in Section \ref{sec:compared_methods} is now done in terms of  computational time. All algorithms were run on a server with two quad core Intel Xeon E5450 processor at 3.00 GHz, 12 MB cache and 16 GB of RAM under MATLAB2014b.\\
Table \ref{tab:time} reports the median, the 5th and 95th percentiles of the computational time over the 200 systems of scenarios S1, S2 and S3, showing a clear gap in the performance of   off-the-shelf methods (PEM, N4SID and SS) and non-off-the-shelf ones (SH, NN, RNN and N2SID); among the latters, our algorithm appears to be the least demanding one.

\begin{table*}
\centering
\scriptsize
{\color{black}
\caption{Computational time (in sec) required to estimate a system: median, 5th and 95th percentiles over 200 Monte-Carlo runs. Estimators are computed using $N_{\cdot,3}=1000$ data (best values among the realistic methods are highlighted in bold).}\label{tab:time}
\begin{tabular}{l|ccc|ccc|ccc}
\toprule
& \multicolumn{3}{c|}{S1} & \multicolumn{3}{c|}{S2}  & \multicolumn{3}{c}{S3} \\
& md & 5th pctl & 95th pctl & md & 5th pctl & 95th pctl & md & 5th pctl & 95th pctl \\
\midrule
SH & 84.89 & 43.70 & 175.24 & 67.22 & 37.49 & 548.89 & 276.62 & 129.07 & 775.38 \\ [-1.5ex]
RNN+CV & 418.93 & 206.28 & 1287.55 & 95.87 & 68.84 & 584.58 & 285.70 & 196.60 & 615.72 \\[-1.5ex]
NN+CV &63.72 & 58.29 & 69.50 & 49.51 & 39.46 & 206.27 & 132.90 & 110.32 & 193.97 \\[-1.5ex]
PEM &3.12 & 2.44 & 4.72 & 1.60 & \textbf{0.70} & 12.89 & 11.47 & \textbf{1.01} & \textbf{31.95} \\[-1.5ex]
N4SID(OE) & \textbf{1.54} & 1.48 & \textbf{1.67} & \textbf{1.46} & 0.96 & \textbf{8.61} & \textbf{7.99} & 1.82 & 36.34 \\[-1.5ex]
SS &1.64 & \textbf{1.47} & 1.84 & 10.51 & 8.86 & 13.23 & 31.33 & 25.58 & 44.09 \\[-1.5ex]
N2SID &666.74 & 508.86 & 851.72 & 576.04 & 462.73 & 732.81 & 504.84 & 402.18 & 764.83 \\
\bottomrule
\end{tabular}
}
\end{table*}

In Section \ref{sec:sgp} a tailored Scaled Gradient Projection (SGP) method has been illustrated to solve the Marginal Likelihood maximization problem at step \ref{alg:ml_max_k} of Algorithm \ref{alg:ident} (see also \eqref{equ:ml_max_opt_probl}). To assess the benefits of SGP, we compare two implementations of Algorithm \ref{alg:ident} which solve the above-mentioned optimization problem using, respectively, the MATLAB routine \verb!fmincon! and  the SGP Algorithm \ref{alg:sgp}. In Table \ref{tab:sgp_fmincon} execution times are reported for the  three scenarios described in Section \ref{subsec:monte-carlo}.
) 
}
\\The routine \verb!fmincon! uses the interior-point algorithm and the default parameters setting (similar performance have been obtained through other algorithms, such as SQP or trust-region-reflective). The parameters involved in the SGP algorithm are set as follows: $\upsilon=10^{-4}$, $\gamma=0.4$, $\alpha_{min}=10^{-7}$, $\alpha_{max}=10^2$, $L_{min}=10^{-5}$, $L_{max}=10^{10}$. The following stopping criterion is adopted:
\begin{equation*}
f(\lambda^{(k)})-f(\lambda^{(k+1)})< 10^{-9} \vert f(\lambda^{(k+1)})\vert
\end{equation*}
For both the algorithms the maximum number of iterations has been fixed to 5000.

\begin{table*}
\centering
\scriptsize
{\color{black}
\caption{Computational time (in sec) required to estimate a system: median, 5th and 95th percentiles over 200 Monte-Carlo runs. Estimators are computed using $N_{\cdot,3}=1000$ data.}\label{tab:sgp_fmincon}
\begin{tabular}{l|ccc|ccc|ccc}
\toprule
& \multicolumn{3}{c|}{S1} & \multicolumn{3}{c|}{S2}  & \multicolumn{3}{c}{S3} \\
& md & 5th pctl & 95th pctl & md & 5th pctl & 95th pctl & md & 5th pctl & 95th pctl \\
\midrule
\verb!fmincon! & 1358.30 & 853.80 & 1893.10 & 2545.10 & 1322.80 & 4816.80 & 6651.60 & 2951.60 & 12732.00 \\
SGP & 84.89 & 43.70 & 175.24 & 67.22 & 37.49 & 548.89 & 276.62 & 129.07 & 775.38 \\
\bottomrule
\end{tabular}
}
\end{table*}

\section{Conclusion} \label{sec:conclusion}
Casting  linear system identification into the Bayesian estimation framework, we have proposed a new  Gaussian prior which has been derived using Maximum Entropy arguments under stability and complexity  (McMillan degree) constraints.  In particular, the part of the prior accounting for complexity controls the rank of the block Hankel matrix built with the Markov coefficients by inducing sparsity on the Hankel singular values; this, in turn, favours the estimated impulse response to lie on what we call ``signal'' subspace, i.e. the subspace spanned by the singular vectors corresponding to the ``non-zero'' Hankel singular values.

We have designed an algorithm which iteratively refines the impulse response estimate by updating the hyper-parameters that define the prior and, in turn, by refining the estimate of the so-called ``signal'' subspace. At each iteration, the main computational burden is given by the hyper-parameters update, which is performed through marginal likelihood maximization. To reduce the computational effort required by this step, a suitably designed Scaled Gradient Projection (SGP) algorithm has been adopted. Simulations have highlighted the significant improvement achieved in terms of execution time of SGP w.r.t. standard off-the-shelf routines.

{\color{black} The numerical comparison illustrated in this paper highlights some advantages of the proposed identification algorithm over state-of-the art routines. First, when MIMO systems have to be identified, our Hankel-based method appears more effective than the original regularization/Bayesian approach relying only on the ``Stable-spline'' kernel. Second, when compared with other methods which include a Hankel-type penalty (such as the the Reweighted Nuclear Norm (RNN)), it provides comparable performance on randomly generated ``large'' MIMO systems, while it appears preferable on a fourth order ``mildly-resonant'' system. 
Third, with respect to more classical approaches, such as PEM and subspace algorithms (N4SID), our method provides more accurate estimates, especially in presence of a small identification dataset.
} 

{\color{black} The analysis of the estimated Hankel singular values has revealed how the final model estimates produced by the proposed algorithm are close to being of ``low order''. Thus, future work will include the design and analysis of tailored model reduction techniques (preliminary work can be found in \cite{PCCDC2015}). Furthermore, we plan to design a more efficient numerical implementation of our algorithm, as well as to extend its application and its comparison with the other routines to the identification of ARMAX models. Finally, a deeper statistical analysis of our approach deserves to be conducted.}



\appendix
\section{Connection with Nuclear Norm minimization approaches}\label{app:connection_nn}
\noindent{\color{black} As observed in \eqref{equ:approx_nn_penalty}, through a special choice of the hyper-parameters $\zeta$ in \eqref{equ:hankel_hyperp}, kernel \eqref{equ:hankel_kernel} induces a nuclear norm penalty on the (squared) Hankel matrix. Previous works in the system identification literature have considered this kind of regularization, starting from the seminal work  \cite{Fazel01}, where the nuclear norm heuristic was proposed for minimal order system approximation.} In the context of subspace-type algorithms,  \cite{LiuV2009} have replaced the SVD step of suitable ``data matrices'' with a   nuclear norm penalty. This approach has then been extended to the case of missing input and output data \cite{LiuHV13} or to short data records \cite{VerhaegenH2014}. Other variations of the method include a nuclear norm weighting \cite{Hansson12} or a nuclear norm minimization algorithm based on reweighting \cite{mohan2010reweighted}. In \cite{GrossmannJM09} similar approaches have been proposed for 
handling missing data scenarios.\\
The approach we propose differs from those discussed above  mainly for three reasons. First, a special weighting scheme, depending upon three hyper-parameters is proposed, which is robust against overfitting and reduces bias. Second, casting the nuclear norm minimization step into a Bayesian framework allows  to use marginal likelihood approaches to estimate the hyper-parameters: 
while these techniques have been shown to be robust
against noise \cite{PCAuto2015}, they also allow to combine the weighted nuclear norm penalty  with other penalties (as we have done in \eqref{equ:ss_hankel_kernel}). Third, while the above-mentioned works adopt a nuclear norm penalty on the Hankel matrix, here the penalty is imposed on the squared Hankel matrix, thus leading to an $\ell_2$ penalty on the Hankel singular values. This is essential in order to derive a Gaussian prior, implying that the marginal likelihood is available in closed form. This facilitates using the marginal likelihood to estimate the hyper-parameters. However, we should stress that in our approach sparsity in the Hankel singular values is favoured by the weighting $\widehat{Q}(\zeta)$.
%

\section{Connection with Iterative reweighted algorithms} \label{app:connection_irw}
\noindent Algorithm \ref{alg:ident} shares key properties with the so-called iterative reweighted algorithms, proposed by \cite{Mohan12jmlr} and \cite{WipfN10}. Considering a rank minimization problem, the algorithm introduced in \cite{Mohan12jmlr} adopts a weighted trace heuristic as a surrogate to the rank function  and iteratively updates the weighting matrix by means of a closed form expression depending on the current optimal point. The trace heuristic considered in \cite{Mohan12jmlr} has a clear analogy to the penalty term \eqref{equ:trace_penalty}, in which $\widehat{Q}(\zeta)$ plays the role of a weighting matrix. Also the  structure of the matrix $\widehat{Q}(\zeta)$ in \eqref{equ:reg:matrix} resembles that of the weighting matrix in \cite{Mohan12jmlr}. Specifically, following the approach in \cite{Mohan12jmlr}, the weighting $\widehat{Q}^{(k)}$ at iteration $k$  would be 
\begin{align}
\widehat{Q}^{(k)}&=\left(\widetilde{\mathcal{H}}(\widehat{\mathbf{h}}^{(k-1)})\widetilde{\mathcal{H}}^\top(\widehat{\mathbf{h}}^{(k-1)}) + \varepsilon I_{pr} \right)^{-1}\label{equ:Q_reweighting}\\
&=\left(\widehat{U}\widehat{S}\widehat{U}^\top + \varepsilon I_{pr}\right)^{-1} \nonumber
\end{align}
where $\widehat{S}$ denotes the singular values matrix and $\varepsilon$ is the regularization factor introduced in order to avoid numerical issues in the matrix inversion operation. Instead, our choice is
\begin{align} \label{equ:Q_nostra}
\widehat{Q}^{(k)}(\hat{\zeta})&= \left(\frac{1}{\hat{\lambda}_1} \widehat{U}_{\bar{n}}\widehat{U}_{\bar{n}}^\top + \frac{1}{\hat{\lambda}_2} \widehat{U}_{\bar{n}}^\perp \left(\widehat{U}_{\bar{n}}^\perp\right)^\top\right)^{-1} \\ 
& = \left(\left(\frac{1}{\hat{\lambda}_1} - \frac{1}{\hat{\lambda}_2}\right)  \widehat{U}_{\bar{n}}\widehat{U}_{\bar{n}}^\top + \frac{1}{\hat{\lambda}_2} I_{pr}\right)^{-1} \nonumber
\\\nonumber
\bar{n}&:=\hat{n}^{(k-1)}
\end{align}
The similarity between \eqref{equ:Q_reweighting} and \eqref{equ:Q_nostra} is apparent with $1/\hat{\lambda}_2$ playing the role of the regularization parameter $\varepsilon$ and\footnote{Note that, even though no such constrained has been introduced, $\hat{\lambda}_1\leq \hat{\lambda}_2$, so that $ \left(\frac{1}{\hat{\lambda}_1} - \frac{1}{\hat{\lambda}_2}\right)>0$.} $\left(\frac{1}{\hat{\lambda}_1} - \frac{1}{\hat{\lambda}_2}\right)  \widehat{U}_{\bar{n}}\widehat{U}_{\bar{n}}^\top$ being a rescaled and truncated version of $\widehat{U}\widehat{S}\widehat{U}^\top$.
%

This peculiar structure of the weighting matrix, which arises from the maximum-entropy derivation of the prior, acts as an hyper regularizer which helps preventing overfitting; the hierarchical Bayesian model  provides a natural framework based on which 
regularization can be tuned through the choice of $\hat \lambda_1$ and $\hat\lambda_2$ (see line \ref{alg:ml_max_k} of Algorithm \ref{alg:ident}).

The Bayesian framework we adopted also connects our algorithm to the non-separable reweighting scheme proposed in \cite{WipfN10} for solving a Sparse Bayesian Learning (SBL) problem: the algorithm iteratively alternates the computation of the optimal estimate and the closed-form update of the hyper-parameters matrix, as the algorithm we propose. The main difference between the cited algorithms and ours  lies in the special structure of the  weighting $\widehat{Q}(\zeta)$, which makes the weighting $\bar{K}_{SH,\eta}$ dependent on the hyper-parameter vector $\lambda=\left[\lambda_0,\lambda_1,\lambda_2\right]$ and $n$ in a way such that closed form expressions for its update are not available. 



\bibliographystyle{plain}                          
\bibliography{References}

\end{document}